\pdfoutput=1 
\documentclass{pic2012}
\usepackage{subfigure,wrapfig}
\def\runauthor{Robert Hirosky}
\def\shorttitle{Jets@Colliders}
\begin{document}

\title{JET PRODUCTION STUDIES AT COLLIDERS}

\author{Robert Hirosky}

\address{University of Virginia\\
Charlottesville, VA\\
E-mail: Hirosky@Virginia.Edu }

\maketitle

\abstracts{An overview of jet production, measurement techniques,
and recent physics results from colliders is presented.  Analyses utilizing
jets and boson plus jets final states are included and implications of the
data are discussed. The results presented here are a snapshot of those
available at the time of the PIC 2012 conference in September 2012.}

\section{Introduction} 
\begin{figure}[!thb]
\begin{center}
\includegraphics[width=0.6\textwidth]{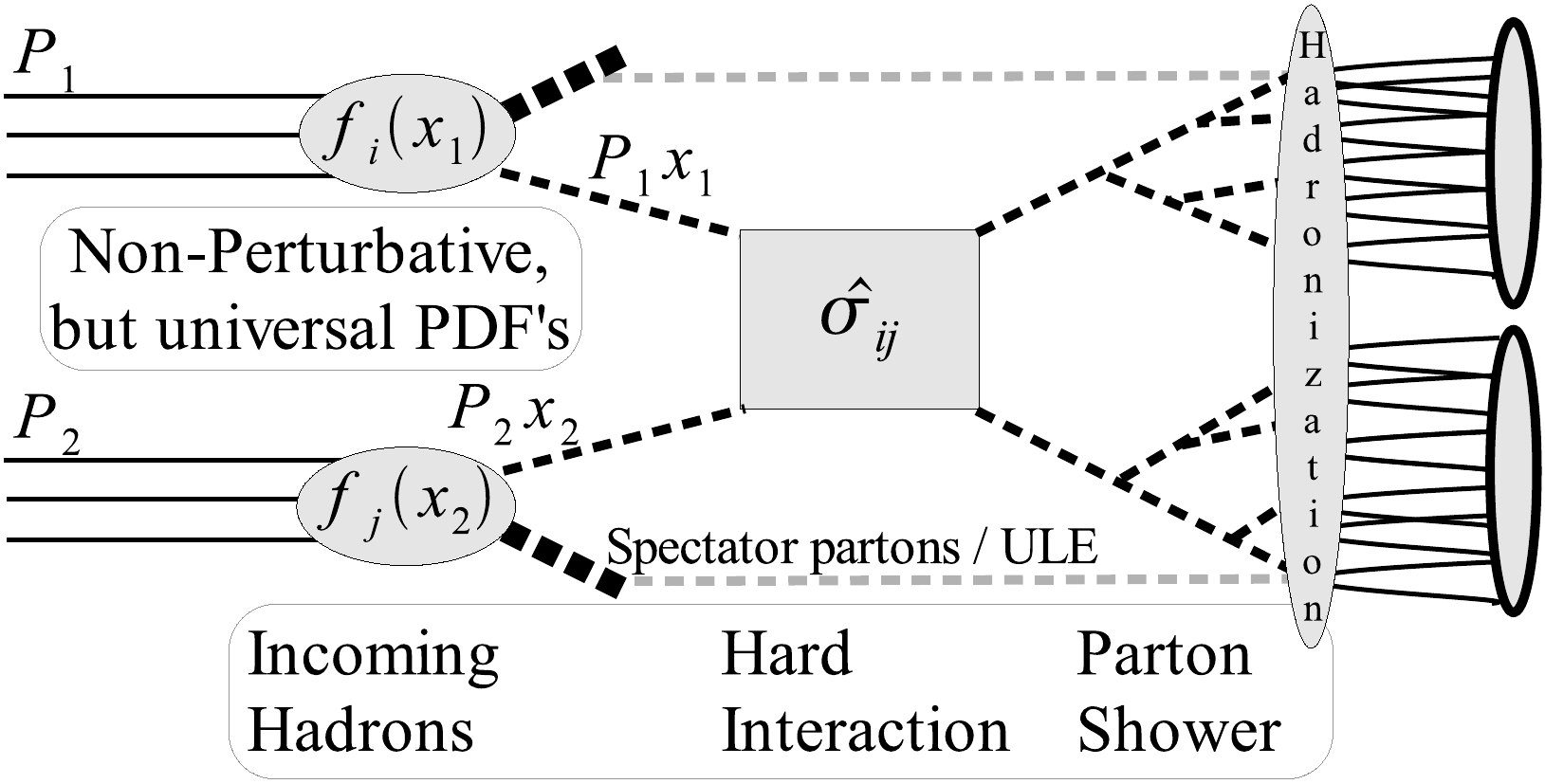}
\caption[*]{ 
Jet production in hadron collisions.  Matrix elements for the 
hard scatter are available at (N)NLO/(N)NLL for many 
processes under study. Non-perturbative parton distribution 
functions and final state hadronization are factorized from the 
hard scattering cross section in calculating production rates at colliders.
}
\end{center}
\label{f:jet_production}
\end{figure}

Hadronic jets are a key signature of strong interactions between
constituent partons in high energy hadron collisions.  The framework of
perturbative quantum chromodynamics (pQCD) describes the partonic cross 
sections~\cite{pQCD} for hard scattering at large momentum transfers
and we have witnessed substantial progress in both experimental and 
theoretical understanding of such processes throughout the past two 
decades.  The production amplitudes for various final states depend on
the convolution of parton level cross sections with experimentally-determined
parton distribution functions (PDFs)~\cite{PDF}.
Pictorially jet production in hadron collisions can be modeled 
as in Fig.~\ref{f:jet_production}.  Matrix elements (MEs) for the hard
interaction are available to (next-to-)next-to-leading-order, (N)NLO,
(next-to-)next-to-leading-logarithm, (N)NLL, for many processes and 
phenomenological models tuned to data can be used to account for
hadronization effects.

Jet production processes include a wide range of phenomena.  In addition to
purely partonic final states, a mix of strong, weak, and electromagnetic 
vertices may be present.  These final states are used to study diverse
topics such as, the performance of pQCD, strong dynamics and couplings, 
hadron structure and compositeness scales, hadronic decay modes of 
short-lived particles, signatures for new massive states, etc.  Furthermore,
the precise knowledge of standard model production rates and kinematic 
properties in jet final states is a prerequisite for achieving highest 
sensitivity in many searches for new physics and rare standard model (SM)
processes.  This paper reviews the status of recent QCD studies from collider
experiments at the LHC and Tevatron.  Discussions of jets in deep inelastic 
scattering, photoproduction, and $e^+e^-$ collisions are included in the 
contributions by M.~Wang and R. Kogler to these proceedings.  An overview
of the status of modern PDF models is presented in the contribution by
Johannes Bl\"{u}mlein.

\section{Jet Identification}
Although hadronic jets arise from the hadronization of elementary quarks 
and gluons and are highly correlated with the four-vectors of their parent 
particle, it is important to note that, unlike elementary particles, 
jets are ultimately defined by the algorithm used in their reconstruction.
Therefore, different algorithms will in general identify a different set of 
jets in the same data.  A good jet algorithm will satisfy the following 
properties: (1) deliver consistent results when applied to partons, particles
from hadronization, or to detector-level information such as tracks or 
energy clusters; (2) be relatively stable with respect to detector noise and
additional energy deposits from hadron remnants, coincidental soft collisions, 
and soft parton radiation; (3) have good energy and angular resolution and be 
relatively straightforward to calibrate.  Typical approaches to jet 
reconstruction employ cone-based or recombinant algorithms.  An extensive review
of modern jet algorithms can be found in Ref.~\cite{fastjet} and 
references therein.

In cone-based algorithms fixed cones with verticies set to the primary 
interaction point are used to select objects to form the jet. 
Typically the angular position of the cone is iterated until its 
geometric center (in azimuthal angle $\phi$ and rapidity $y$) 
matches the location of the
combined four-vector or weighted average position of the enclosed objects
(energy deposits, tracks, particle four-vectors, etc).
The cone algorithms primarily used at the Tevatron are described in 
Refs.~\cite{D0Cone} and~\cite{CDFCone} for the D0 and CDF Experiments, 
respectively.

Recombination or sequential clustering algorithms successively merge 
objects based on spatial and/or
relative transverse momenta criteria.  Clusters are then successively merged
until satisfying some well defined stopping criteria and the jet four-vector
is calculated by summing all combined objects.  Examples are the 
$k_T$~\cite{kt} algorithms and also the anti-$k_T$ algorithm~\cite{antikt} 
which is frequently employed in analyses from the LHC.

While various experiments and individual analyses can employ different
algorithms or choices of parameters to control their performance, 
each measurement is necessarily compared to theory or Monte Carlo 
(MC) using consistent algorithm definitions.

\section{Inclusive Jets and Dijet Production}

Measures of inclusive jet production are sensitive to a combination of
QCD matrix elements describing the hard parton scattering and initial state
parton distribution functions (PDFs).  Inclusive measures allow tests of
perturbative QCD over wide ranges of parton momentum exchange ($Q$) and 
their sensitivity to wide ranges of parton momentum fractions ($x$) can 
provide further constraints to PDF models.  Furthermore, because the strong 
coupling ($\alpha_s$) depends on the momentum exchange scale, 
these measurements provide information about the running of the coupling 
to our largest accessible momentum scales.

\begin{figure}[!thb]
\begin{center}
\subfigure[Inclusive jet cross section measurements by the 
D0 Experiment as a function of jet $p_T$ in six $|y|$ bins.]{%
            \label{f:d0incl}
            \includegraphics[width=0.46\textwidth]{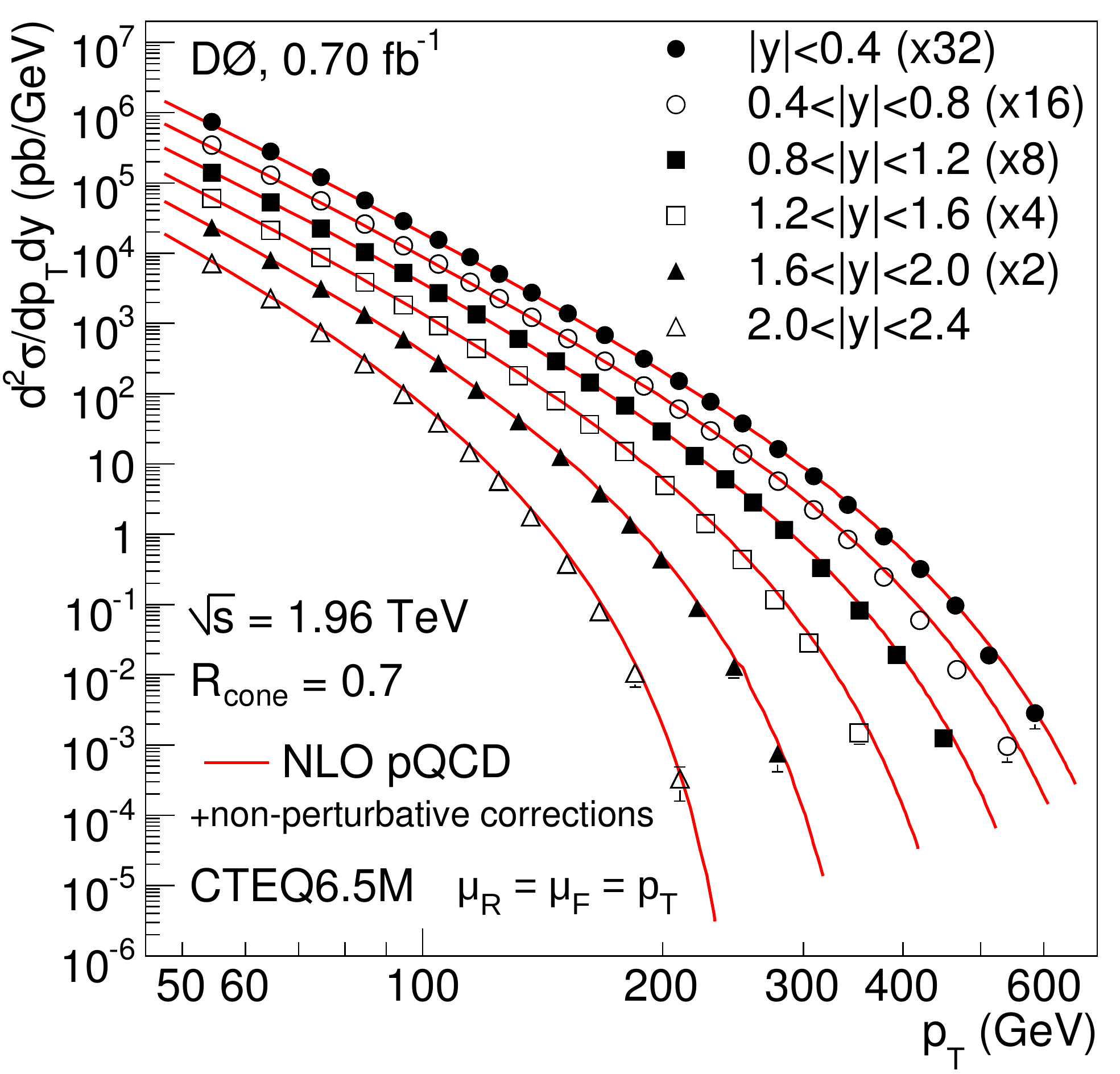}
        }%
\hfill
\subfigure[Inclusive jet cross section measurements by the 
CMS Experiment as a function of jet $p_T$ in five $|y|$ bins.]{%
           \label{f:cmsincl}
           \includegraphics[width=0.46\textwidth]{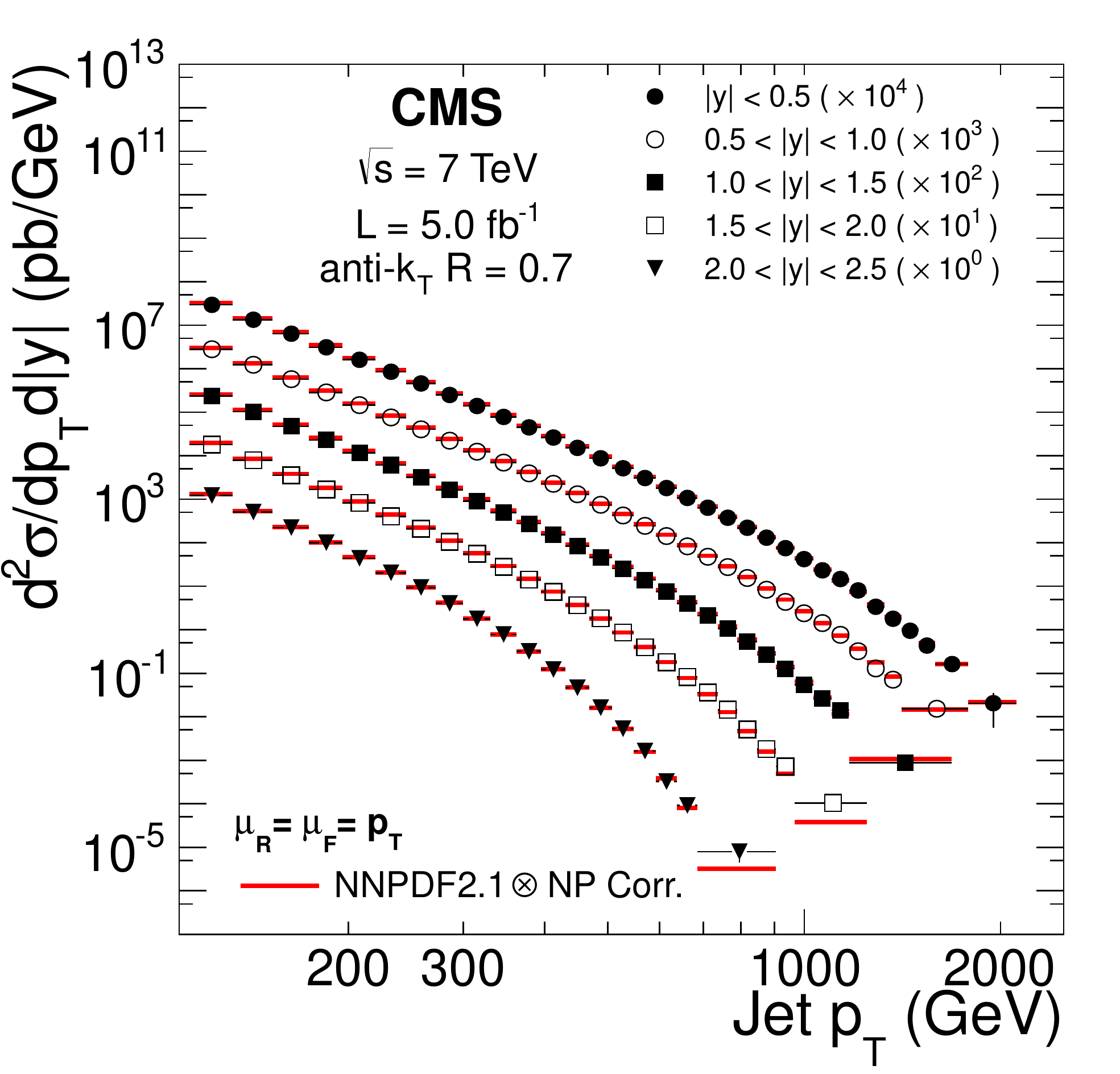}
        }
\caption[*]{Measures of the inclusive jet cross section at the 
Tevatron and LHC.}
\end{center}
\label{f:incl}
\end{figure}

Measurements of the inclusive jet cross section plotted differentially in
bins of jet transverse momentum ($p_T$) and rapidity ($y$) are shown in 
Figs.~\ref{f:d0incl}~and~\ref{f:cmsincl} for the D0~\cite{D0incjets} 
and CMS~\cite{CMSincdijets} Experiments, respectively.  These results 
illustrate the remarkable success of perturbative
QCD and phenomenological understanding of proton structure in 
high energy collisions, spanning 8--9 orders in jet $p_T$.  In each case 
experimental results are compared to theory calculated to 
next-to-leading-order in QCD.

\begin{figure}[!thb]
\begin{center}
\subfigure[ATLAS measurement of dijet double-differential 
cross section plotted versus invariant mass of the leading $p_T$ jets.
Measurements are binned in terms of the average rapidity separation 
between the jets $y^*=|y_1-y_2|/2$.]{%
            \label{f:ATLASdijet}
            \includegraphics[width=0.46\textwidth]{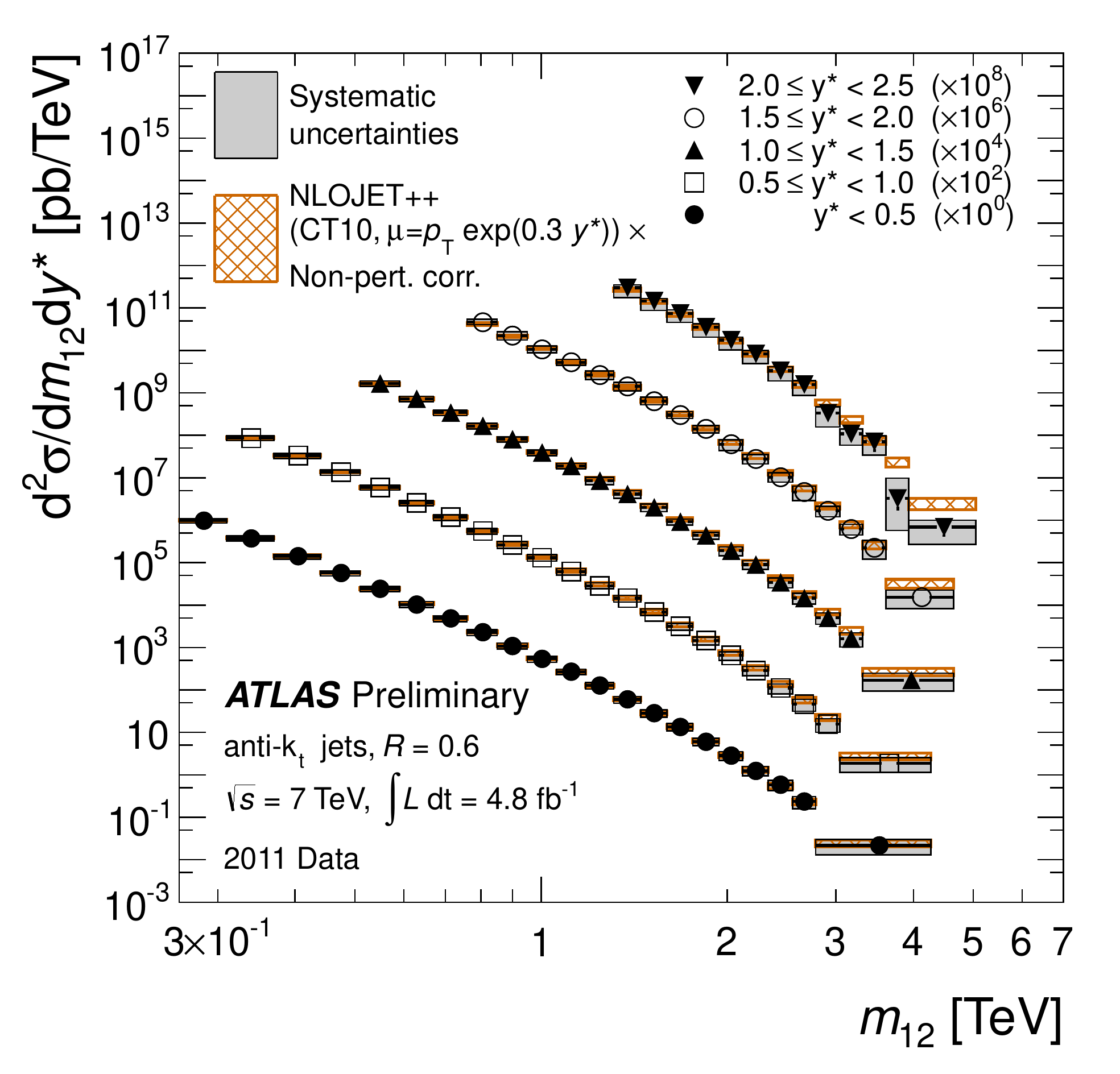}
        }%
\hfill
\subfigure[CMS measurement of the dijet double-dif-ferential
cross section plotted versus invariant mass of the leading $p_T$ jets. 
Measurements are binned in terms of the maximum rapidity of the two 
jets $|y_{max}|={\rm max}(|y_1|,|y_2|)$.]{%
           \label{f:CMSdijet}
           \includegraphics[width=0.46\textwidth]{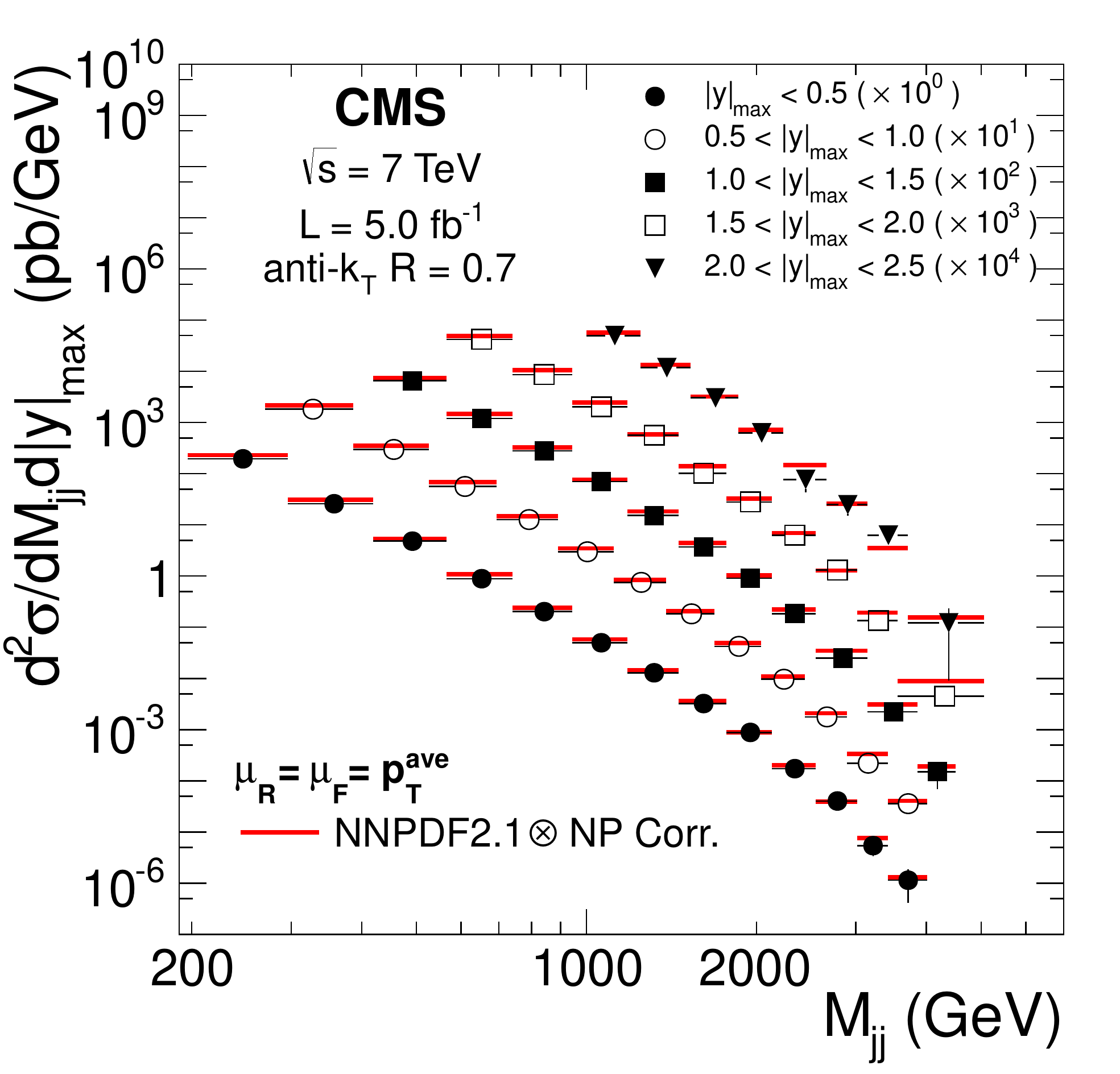}
        }
\caption[*]{Measurements of the dijet mass cross sections at the LHC.}
\end{center}
\label{f:LHCdijet}
\end{figure}

Measurements of the dijet mass spectra in LHC collisions at 7\,TeV
presented by the ATLAS~\cite{ATLASdijet7} and CMS~\cite{CMSincdijets} 
Experiments are shown in Fig.~\ref{f:LHCdijet}. Preliminary 
results at 8\,TeV are in agreement with these~\cite{ATLASdijet8}.  
Again, detailed comparisons to predictions from pQCD show good 
agreement between data and theory.

\begin{figure}[!thb]
\begin{center}
\subfigure[D0 data divided by theory for the inclusive 
jet cross section as a function of jet $p_T$ in six $|y|$ bins.]{%
            \label{f:D0pdf}
            \includegraphics[width=0.55\textwidth]{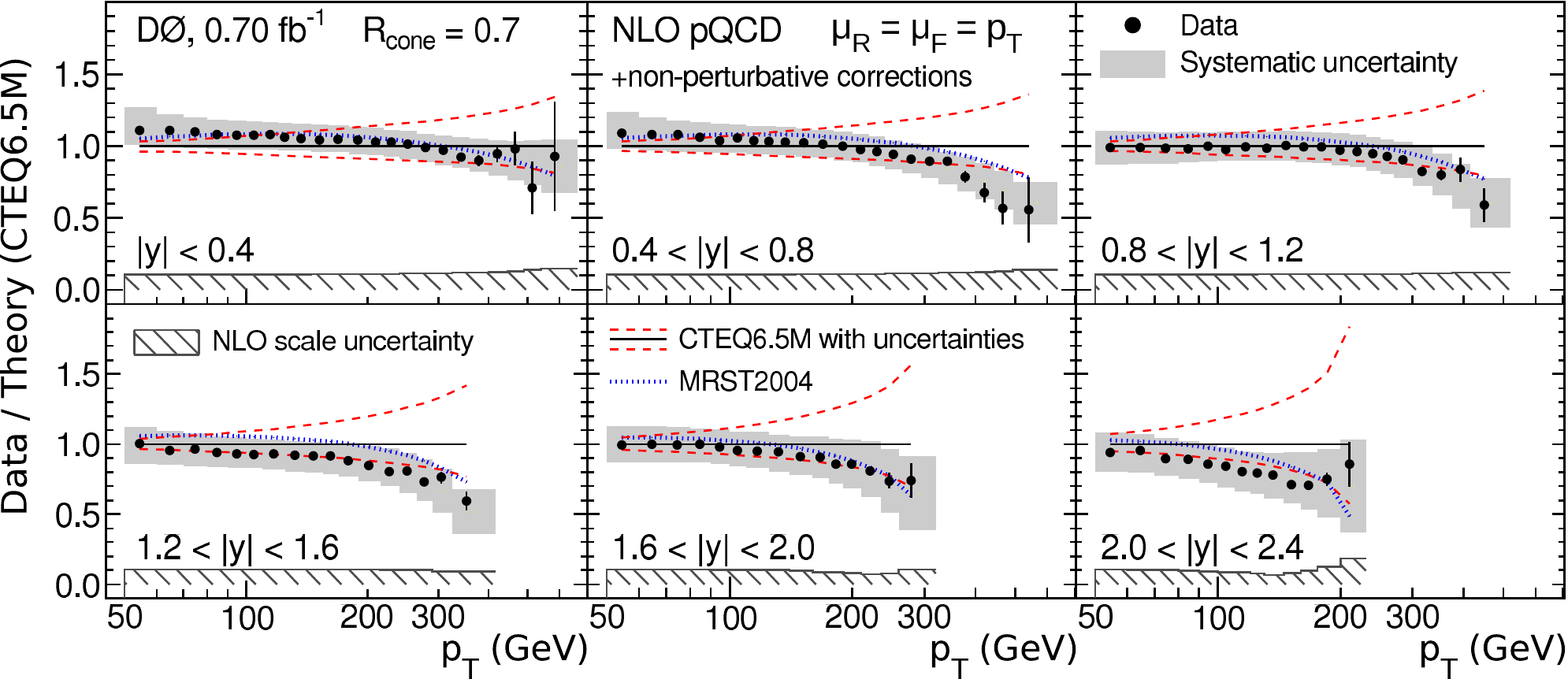}
        }%
\hfill
\subfigure[CMS ratio of inclusive jet cross section ($|y| < 0.5$) to the theory
prediction with NNPDF2.1~\protect\cite{nnpdf} PDFs.]{%
           \label{f:CMSrat}
           \includegraphics[width=0.39\textwidth]{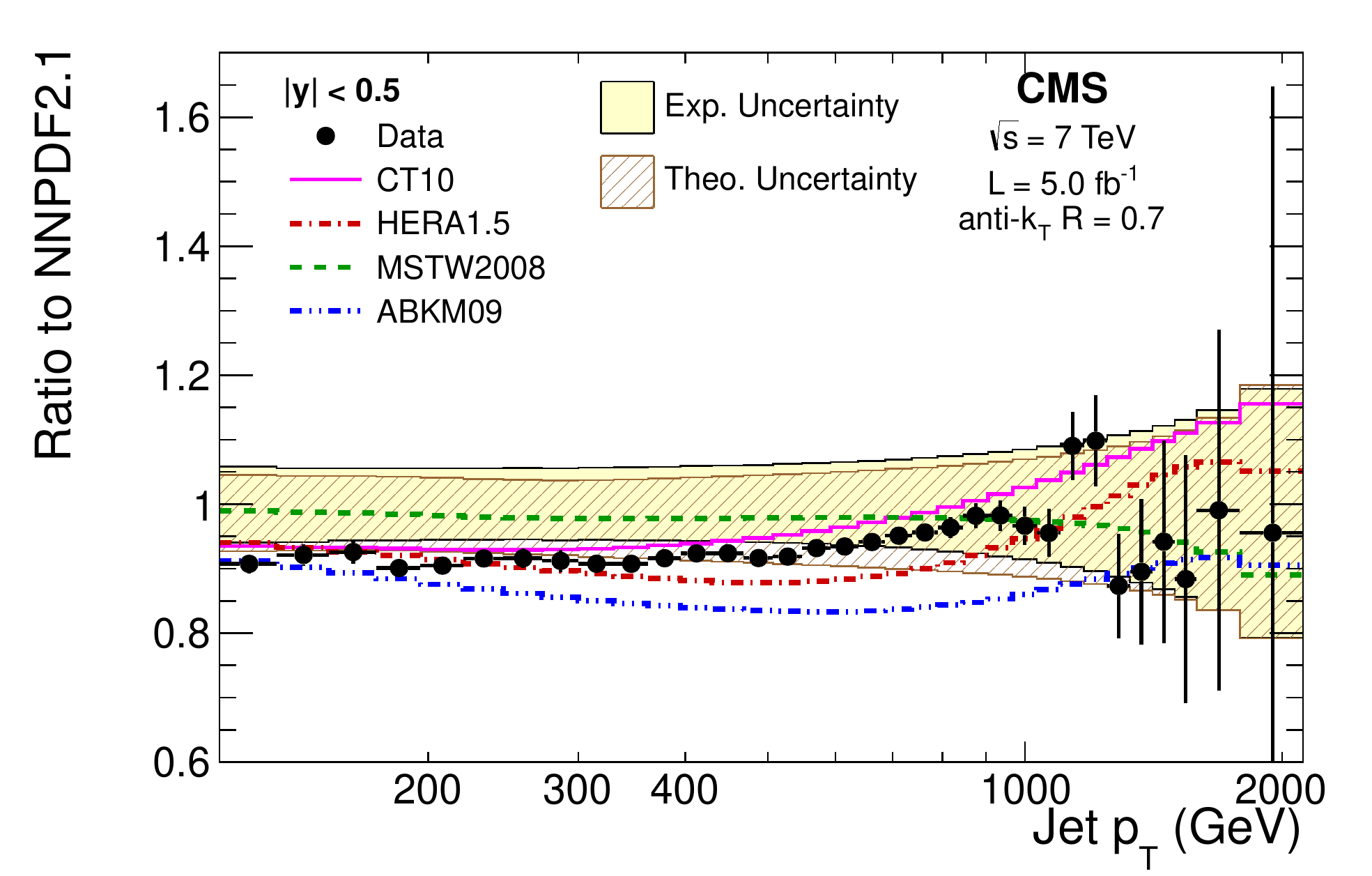}
}%
\caption[*]{Examples of relative precision of experimental measures and theory
calculations on inclusive jet cross sections.}
\end{center}
\label{f:pdf}
\end{figure}

Examples of the relative precision of the inclusive jet measurements are shown 
in Figs.~\ref{f:D0pdf}~and~\ref{f:CMSrat}.  
In these results, illustrating D0 and CMS measurements,
the experimental uncertainties are similar to 
or significantly smaller than those associated
with PDF models, showing that the data can provide significant constraints
on the PDFs.  A similar comparison from the ATLAS Experiment
is shown in Fig~\ref{f:ATLASdijetComp}, but considering
the dijet mass spectrum.  This also illustrates the interest 
of pushing experimental measurements to more extreme regions of phase space for 
jet production.  For dijet pairs at the maximum measured rapidity separation,
the agreement of theory and experiment is clearly observed to break down at
large invariant mass, which may indicate the need for higher order terms to 
accurately describe the data.

\section{Measurements in Multijet Final States}

Measurements using multijet final states are powerful tools to test the 
applicability of pQCD calculations and to examine the running of the 
strong coupling constant.  The D0 Experiment reported an analysis of
multijet final states using a new observable~\cite{D0RDR}, 
$R_{\Delta R}$, which is computed as a ratio of cross sections as 
illustrated in 
Fig.~5. 
This observable measures
the number of neighboring jets that accompany a jet of given $p_T$ 
within an angular region $\Delta R$ defined in $(y,\phi )$ space.
Because PDF dependencies largely cancel in the ratio, these results
can be used to extract a measure of $\alpha_s$, 
almost independent of initial assumptions on the 
renormalization group equation (RGE).
The measured results are displayed in Fig.~\ref{f:RDRXS} in bins of
$\Delta R$, the search region for neighboring jets,
and $p_{Tmin}^{nbr}$, the minimum $p_T$ for inclusion of
neighboring jets as a function of inclusive jet $p_T$.  In each $\Delta R$
interval the value of $R_{\Delta R}$ increases with jet $p_T$ 
until approaching the kinematic limit. 

Event shape studies~\cite{CMSmulti} by the CMS Experiment include a 
measure of the central transverse thrust, $\tau_{\perp,\cal C}$ which
probes QCD radiative processes and is mostly sensitive to the modeling 
of two- and three-jet topologies.  The central transverse thrust
defined in Refs.~\cite{CMSmulti,tauC} is a measure of radiation
along an event's transverse thrust axis.  The CMS measurement is shown in
Fig.~\ref{f:CMSmulti} compared to predictions from five MC simulations.
The parton shower (PS) MC generators {\sc pythia}~\cite{pythia} 
and {\sc herwig}~\cite{herwig}
show generally good agreement with the data, except for {\sc pythia}
version~8
which shows some discrepancy at extreme values of $\tau_{\perp,\cal C}$.
The disagreement in calculations by
{\sc alpgen}~\cite{alpgen} and {\sc MadGraph}~\cite{madgraph} 
implies that the regime of high jet $p_T$ 
and multiplicity where the explicit higher final
state parton multiplicity ME calculations of these generators 
significantly improves upon a pure PS approach has not been reached in 
these data.

\begin{figure}[!thb]
\begin{center}
\includegraphics[width=0.7\textwidth]{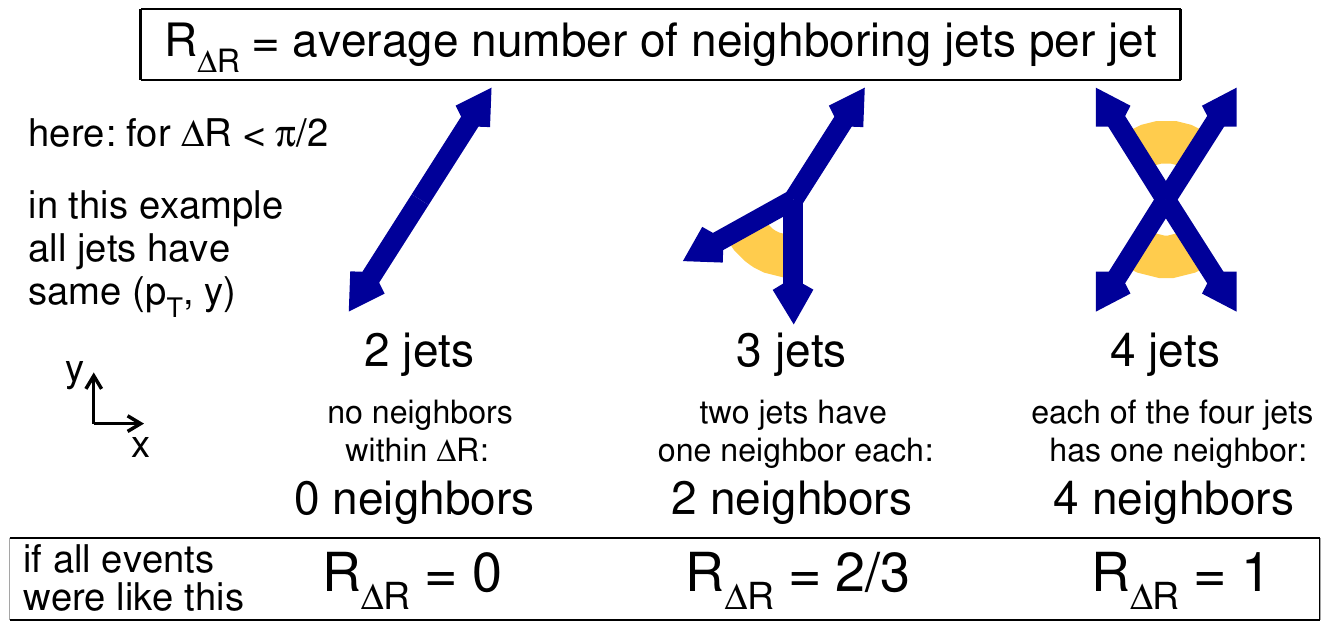}
\caption[*]{Illustration of the variable 
$R_{\Delta R}(p_T,\Delta R,p_{Tmin}^{nbr}) = $ \hfill\mbox{}
$\sum_{i=1}^{N_{jet}(p_T)} N_{nbr}^{(i)}(\Delta R,p_{Tmin}^{nbr}) / N_{jet}(p_T) $, 
which measures the number of neighboring jets accompanying a jet satisfying 
a given requirement on transverse momentum.}
\end{center}
\label{f:RDR}
\end{figure}

\begin{figure}[!thb]
\begin{center}
\subfigure[ATLAS ratios of dijet double-differential cross section, shown
as a function of dijet invariant mass, to the theoretical prediction 
obtained using {\sc NLOJET++}~\protect\cite{nlojet++} with the 
CT10~\protect\cite{ct10} PDF set.]{%
            \label{f:ATLASdijetComp}
            \includegraphics[width=0.5\textwidth]{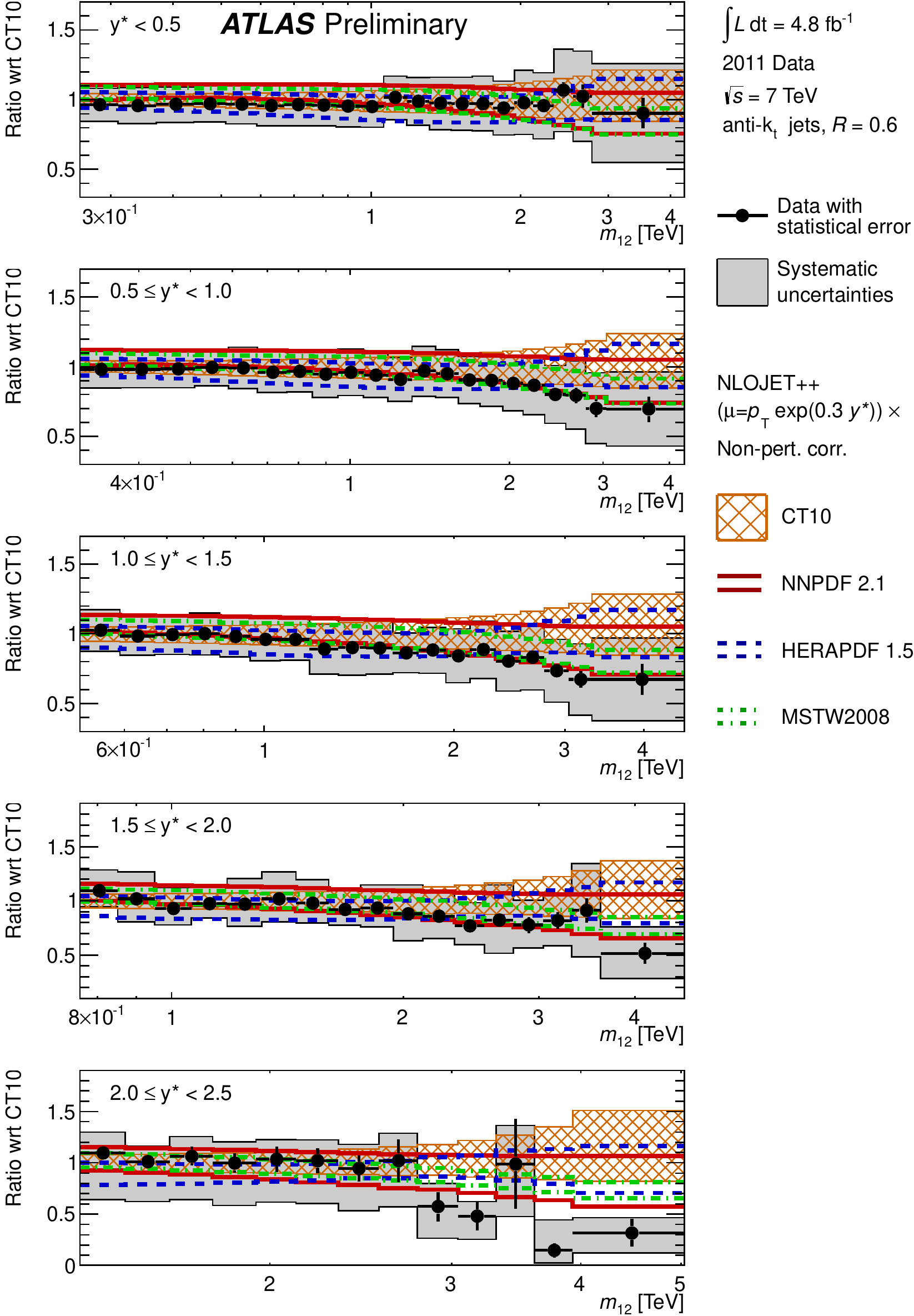}
        }%
\hfill
\subfigure[CMS Distributions of the logarithm of the central 
transverse thrust for events with leading jet between 125 and 250\,GeV.
Results in data are compared to predictions from five MC simulations.
Shaded bands represent the quadratic sum of the statistical and 
systematic uncertainties.]{%
\label{f:CMSmulti}
\includegraphics[width=0.42\textwidth]{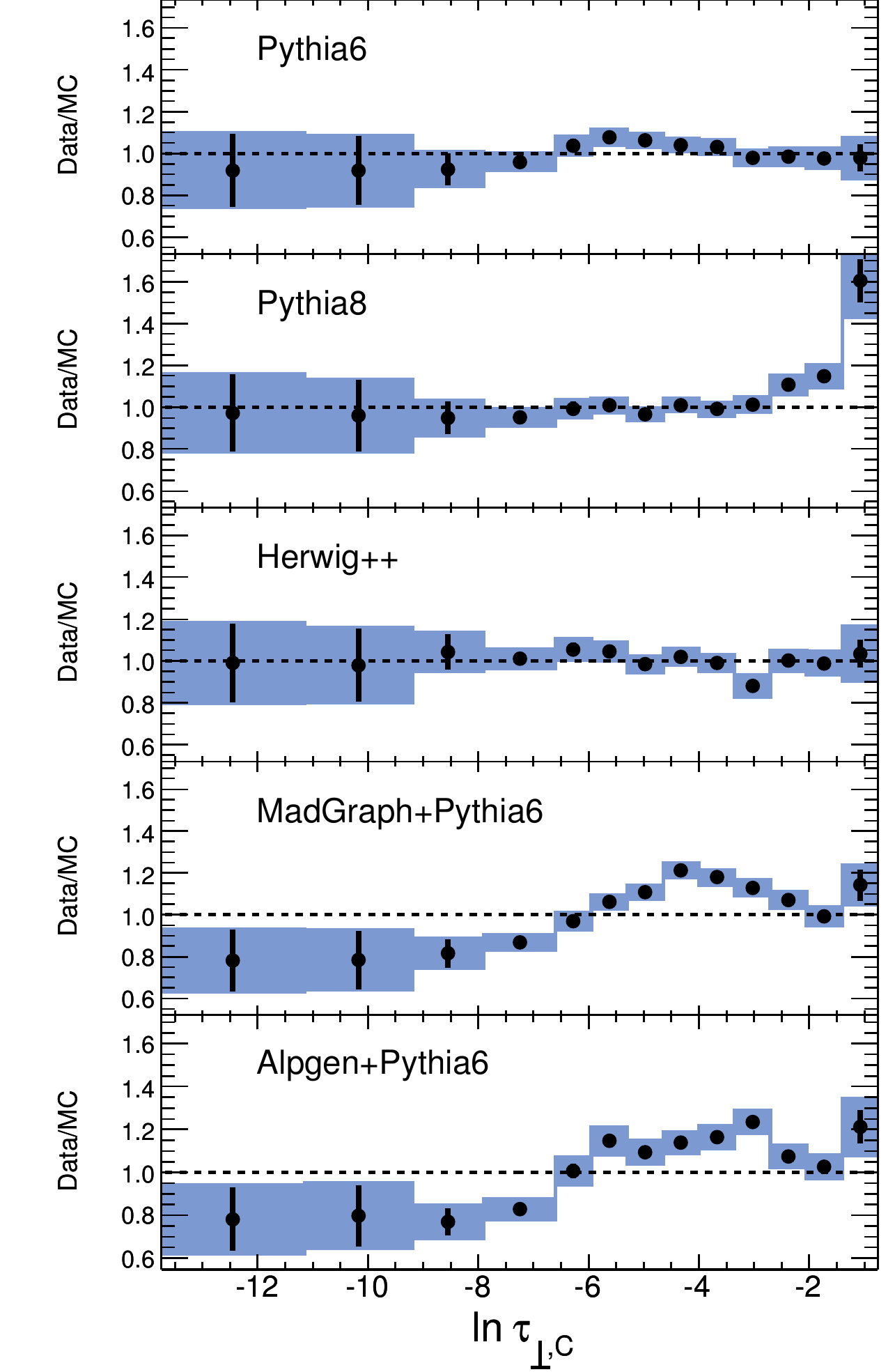}
}
\caption[*]{Measures of dijet and multijet distributions at the LHC.}
\end{center}
\label{f:LHCmulti}
\end{figure}

\begin{figure}[!thb]
\begin{center}
\subfigure[Measurement of $R_{\Delta R}$ for different intervals in
the $\Delta R$ search region for neighboring jets and requirements on
$p_{Tmin}^{nbr}$.]{%
            \label{f:RDRXS}
            \includegraphics[width=0.63\textwidth]{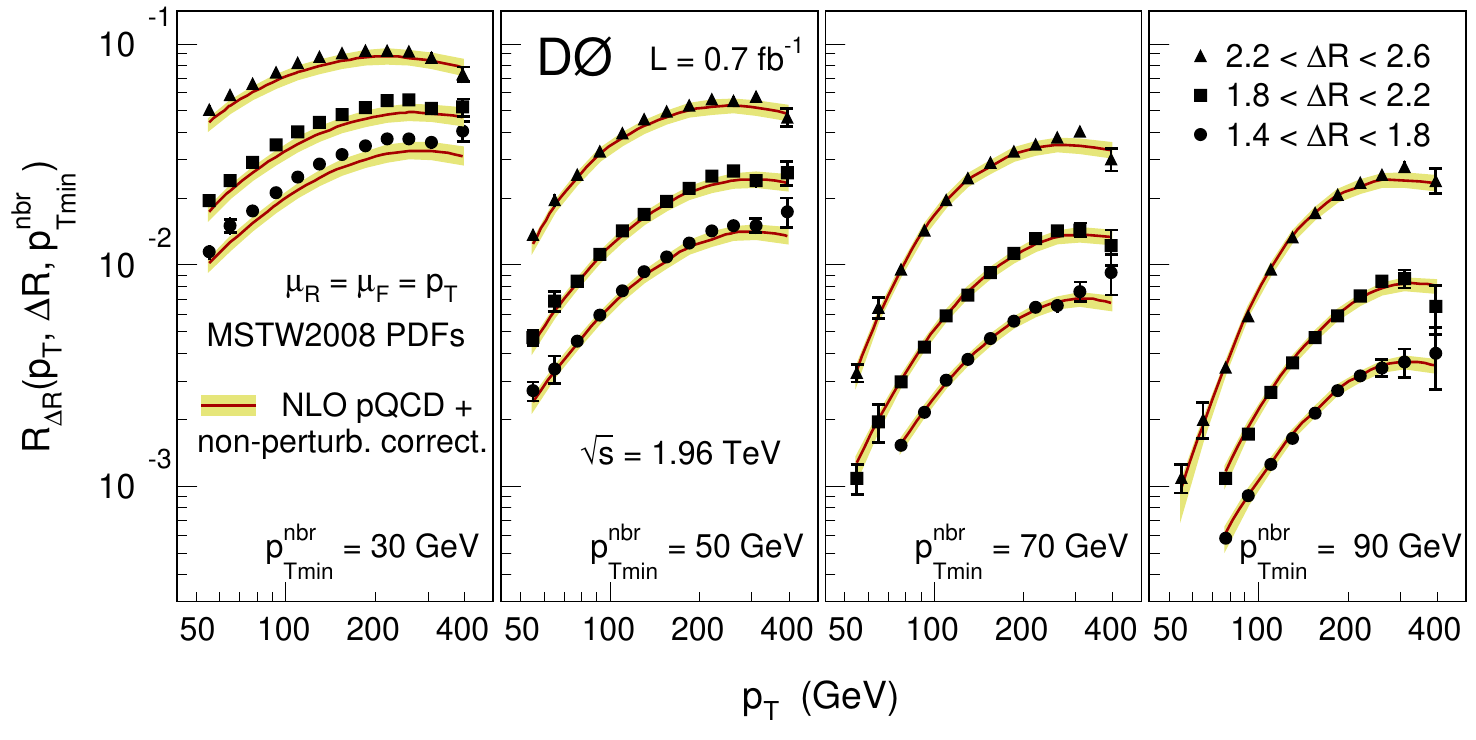}
        }%
\hfill
\subfigure[Measurements of the strong coupling 
constant $\alpha_s$ as a function of 
momentum transfer $Q$. ]{%
           \label{f:D0alphas}
           \includegraphics[width=0.32\textwidth]{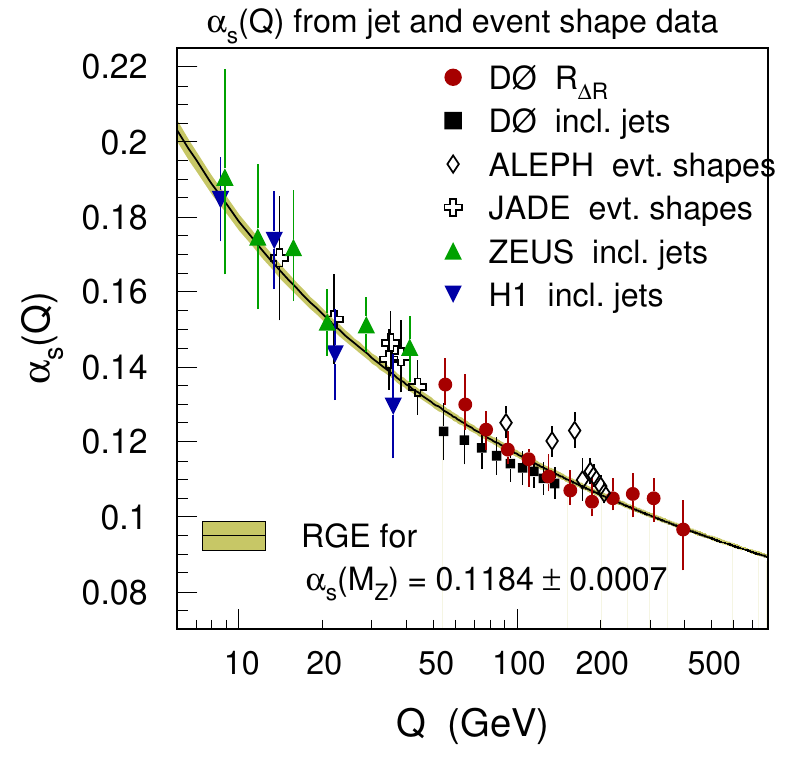}
        }
\caption[*]{D0 measurements of (a) multijet distributions and 
(b) extraction of $\alpha_s$, determined from D0 inclusive jets and event 
shapes.  }
\end{center}
\label{f:D0multi}
\end{figure}

\section{Measure of the Running of the Strong Coupling Constant}
Precise measures of jet production can be used to test the running of the
strong coupling constant $\alpha_s$, reaching significantly higher scales than
are accessible by other methods.  While the running of $\alpha_s$ is predicted
by the RGE, it remains the least precisely
known of the fundamental couplings.  Furthermore, its running can be modified 
at large $Q$ by the presence of new physics, such as the presence of extra
spatial dimensions.

Because jet cross sections depend on a combination the strong coupling and 
parton distributions, the extraction of $\alpha_s$ requires 
theory predictions as a continuous function of $\alpha_s$ used in both 
the matrix elements and PDFs.  The dependence of the PDFs on $\alpha_s$
is parameterized by interpolating theory results based on
PDF sets determined using incremental steps in $\alpha_s$.
Correlated systematic experimental and theoretical uncertainties are 
treated via the Hessian approach~\cite{Hess} and $\alpha_s$ values are
determined by minimizing a $\chi^2$ function with respect to $\alpha_s$
and nuisance parameters for the uncertainties.  Results are show in 
Fig.~\ref{f:D0alphas} for this measurement and an earlier
determination~\cite{D0alphas} using the D0 inclusive jet cross section.
Additional determinations of the strong coupling constant 
using inclusive and multijet cross sections 
in deep inelastic scattering data are presented
in the contribution by R. Kogler to these proceedings.

\section{Jets in Association with Vector Bosons}

A great deal of progress has been made in measurements of jets in association 
with vector bosons ($V$) in recent years.  These processes provide 
valuable tests of 
fixed order and ME+PS predictions and their precise measurements provide 
important constraints on backgrounds for rare SM physics and searches for 
evidence of new physics processes. The production of
$V + n$-jet final states provides a handle to study multiscale QCD processes
and plays a significant role as background in searches for beyond the standard 
model phenomena and many Higgs boson measurement channels. 
Theoretical uncertainties on their production rates and kinematics 
introduce large uncertainties and limit our ability to identify
new physics processes.  

A large variety of Monte Carlo programs are compared to data to test the
applicability of perturbative calculations and phenomenologically tuned
models in diverse regions of phase space for $V + n$-jet final states.
Examples of MC calculations compared to these data include:
LO parton shower MC such as {\sc pythia}, 
leading-order matrix element plus parton shower matched MC such as 
{\sc alpgen} and {\sc sherpa}~\cite{sherpa}, calculations
employing all-order resummation of wide angle emissions 
({\sc hej}~\cite{hej}), and next-to-leading order pQCD predictions
for the production of a vector boson plus multiple parton final states:
{\sc blackhat+sherpa}~\cite{blackhat} and {\sc rocket+mcfm}~\cite{rocket,mcfm}.

\begin{figure}[!thb]
\begin{center}
\subfigure[D0 measurement of the W boson transverse momentum 
distributions in inclusive $W+n$-jet events for $n = 1-4$ 
compared to various theoretical
predictions.]{%
            \label{f:D0Wpt}
            \includegraphics[width=0.46\textwidth]{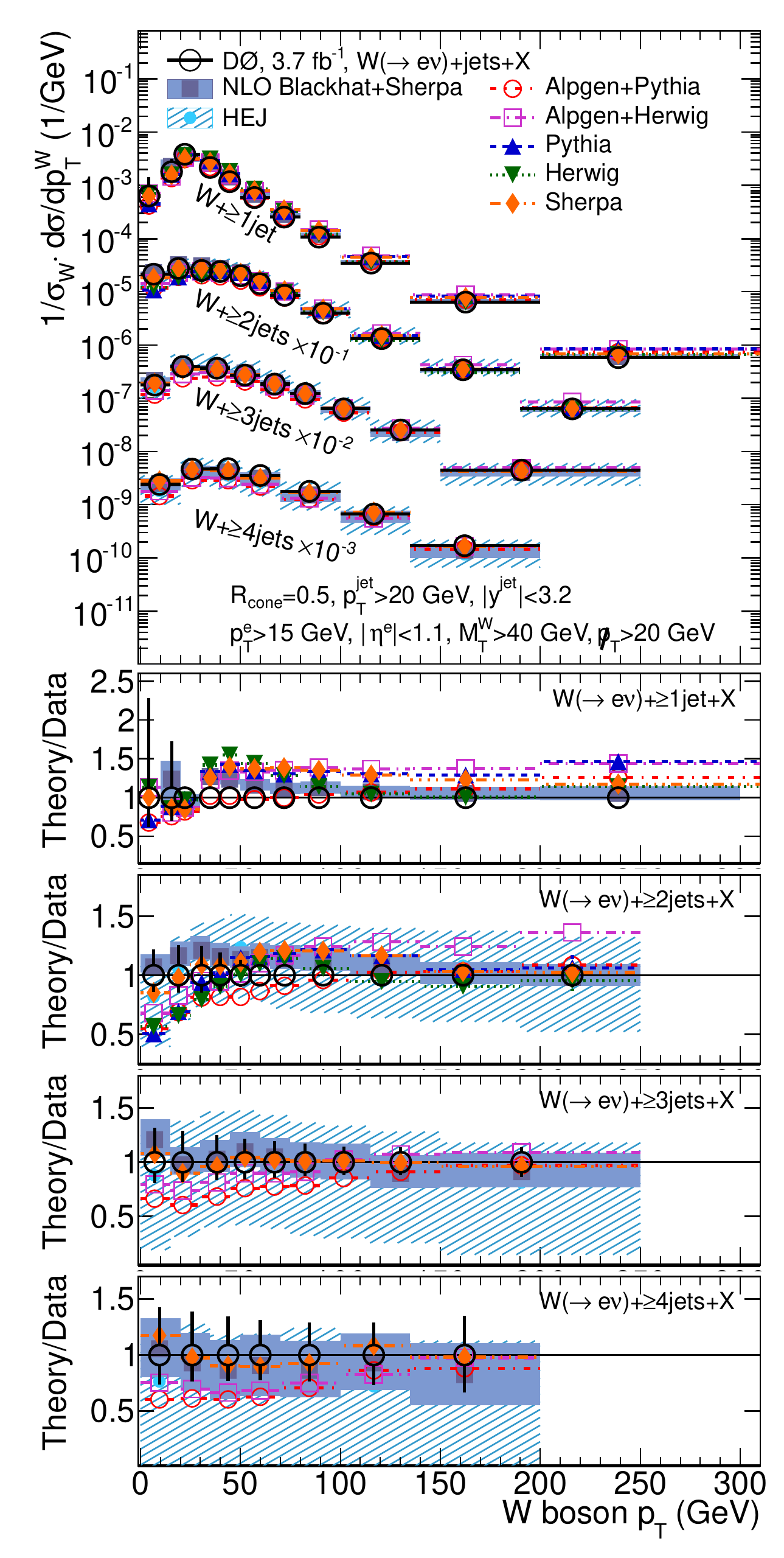}
        }
\hfill
\subfigure[ATLAS W+jets cross section as a function
of $H_T$ for separate jet multiplicities.   Results are
compared to predictions from {\sc alpgen}, {\sc sherpa}, and 
\sc {blackhat-sherpa}.]{%
           \label{f:ATLASWHT}
           \includegraphics[width=0.46\textwidth]{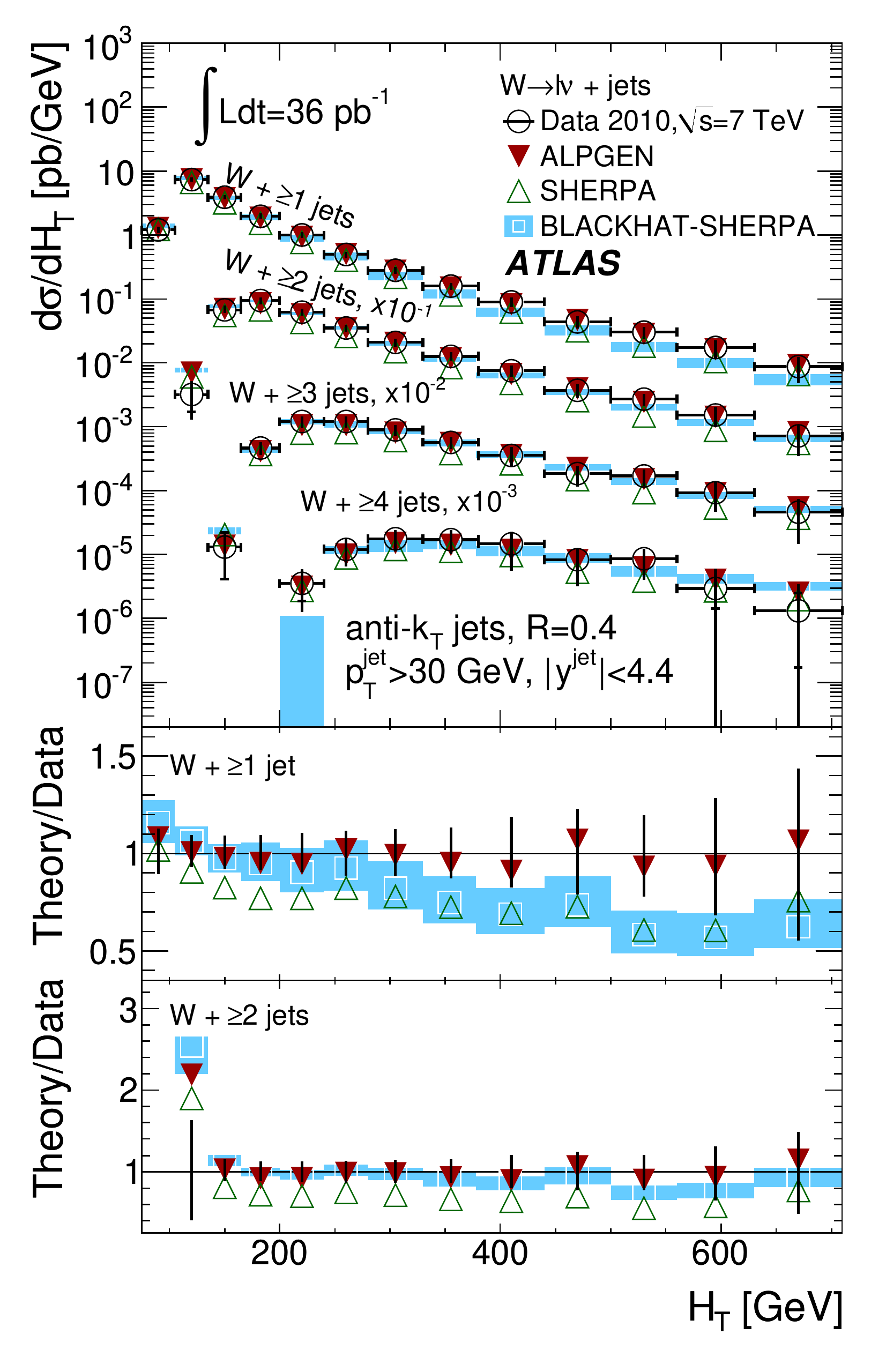}
        }
\caption[*]{ $W+n$-jet measurements from the D0 and ATLAS Experiments.}
\end{center}
\label{f:Wjets}
\end{figure}

Figure~8  
shows examples of measurements performed
at the D0~\cite{D0Wj} and ATLAS~\cite{ATLASWj} Experiments.
The D0 measurement of the $W$ boson transverse momentum for inclusive
jet multiplicity bins is shown in Fig.~\ref{f:D0Wpt}.
Predictions from {\sc blackhat-sherpa} and {\sc hej}
show good agreement for all jet multiplicities.  At 
jet $p_T$ below the threshold of 20\,GeV, nonperturbative 
effects dominate and the fixed order calculations are
expected to be unreliable.

The measured $H_T$ distribution (scalar $p_T$ sum of reconstructed 
physics objects) for $W$ events with one or
more jets, measured at ATLAS, is shown in Fig.~\ref{f:ATLASWHT}.
The prediction, calculated inclusively at NLO
by {\sc blackhat-sherpa}, does not describe the data,
due to the limited order of the matrix
element calculation, which does not include three or more real emissions 
of final-state partons.  The LO ME calculation of {\sc alpgen} with up to
five final-state partons, describes the data well. 
A study described in Ref.~\cite{ATLASWj} considers improvements
with a modified treatment of {\sc blackhat-sherpa} predictions
introducing higher-order NLO terms describing higher real emission 
multiplicities.  A matching scheme is developed and required to 
reduce double-counting 
of cross sections, illustrating the challenges of comparing NLO calculations
to complex inclusive jet variables like $H_T$.

\begin{figure}[!thb]
\begin{center}
\subfigure[D0 measurement of the probability of
emission of a third jet in inclusive $W+2$-jet events for various
definitions of rapidity interval as described in the text.]{%
            \label{f:D0Wem}
            \includegraphics[width=0.46\textwidth]{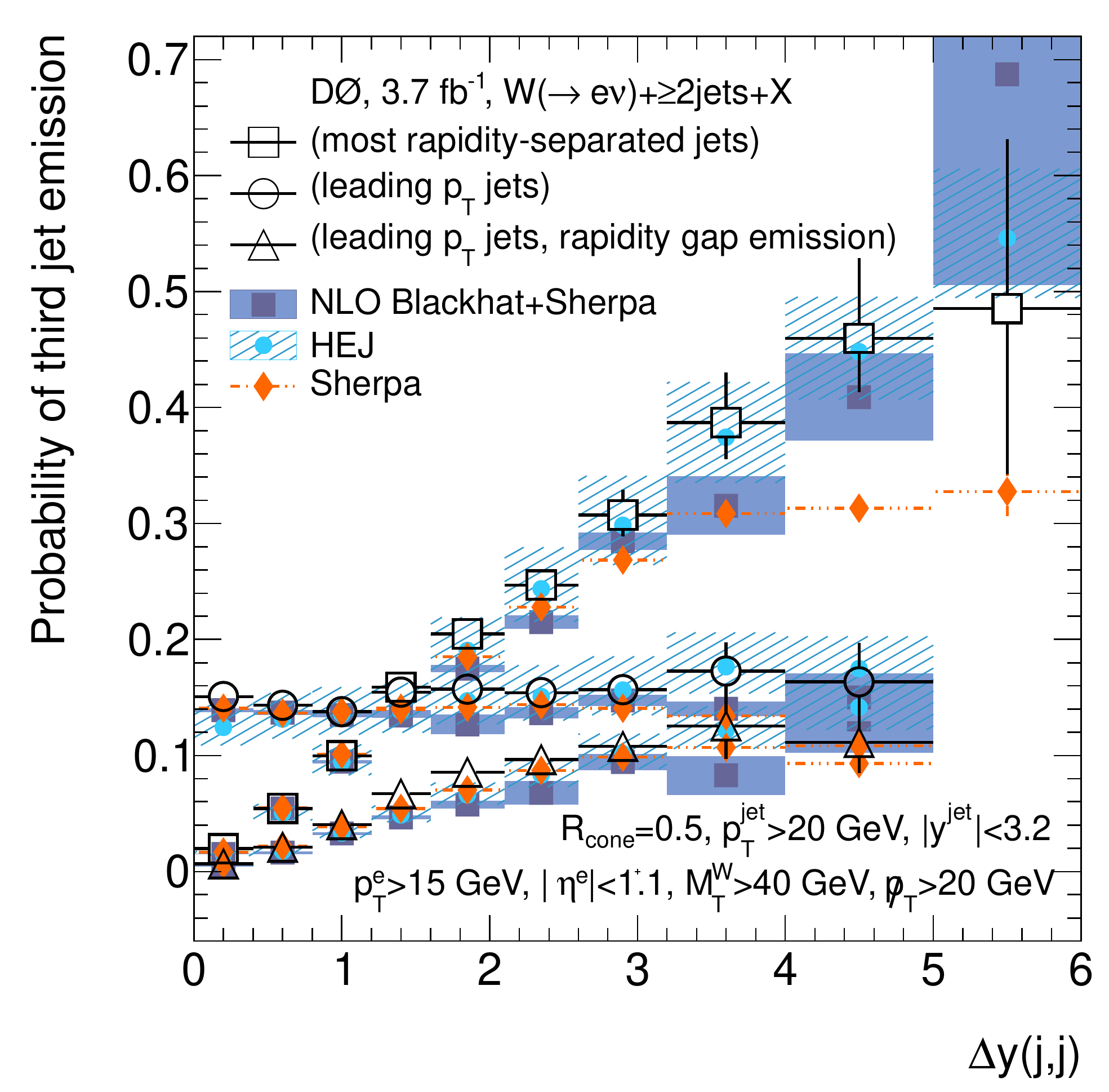}
        }
\hfill
\subfigure[CDF measured cross section in $Z/\gamma^*+\ge 1$ jet events 
as a function of leading lepton $p_T$.]{%
           \label{f:CDFZlpt}
           \includegraphics[width=0.46\textwidth]{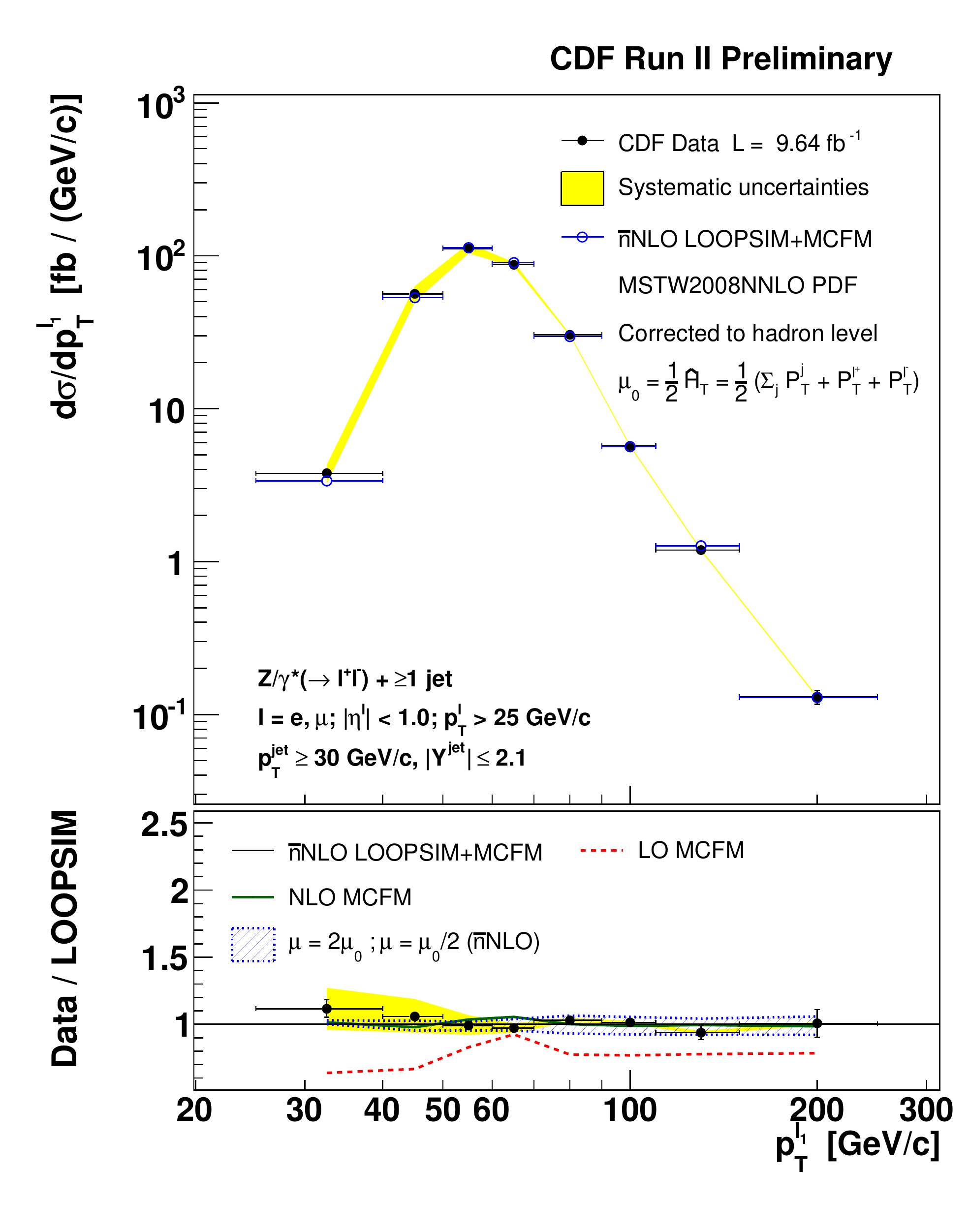}
        }
\caption[*]{Measurements of $W$ and $Z+$jets differential cross sections from
the Tevatron.}
\end{center}
\label{f:WZjets}
\end{figure}

The D0 Experiment reported a first measurement of the probability of
emission of a third jet in inclusive $W+2$-jet events as a 
function of the dijet rapidity separation of the two leading jets.
This measurement, shown in Fig.~\ref{f:D0Wem}, 
is presented in
both pT- and rapidity-ordered scenarios, and a hybrid
result considering the probability of additional jet emission in the
rapidity interval defined by the two leading $p_T$ jets. 
Resummation predictions from {\sc hej} are best able to describe 
both the rate and shape across the full rapidity range.

Studies involving differential properties in $Z+$jet production are 
similarly of interest to those involving $W+$jets.  Using leptonic decay
modes of the $Z$ boson allows for selection of low background
samples and provides precise constraints on the scale of hadronic recoil
in the event.  CDF reported measured $Z+$jet cross sections compared to 
NLO QCD combined with NLO EW calculations~\cite{NLOEW}.  A 
measurement of the $Z+$jet cross section versus 
$p_T$ of the leading lepton from the $Z$ 
decay~\cite{CDFZlpt} is shown to be well modeled in Fig.~\ref{f:CDFZlpt}.

\begin{figure}[!thb]
\begin{center}
\subfigure[CDF fit to selected events in $W+b$-jet analysis, based
on a maximum likelihood of the ratio of vertex mass distributions 
for signal and background in $b$ tagged jets in the selected data sample in
1.9\,fb$^{-1}$ of Tevatron Run 2 data.]{%
            \label{f:CDFWb}
            \includegraphics[width=0.46\textwidth]{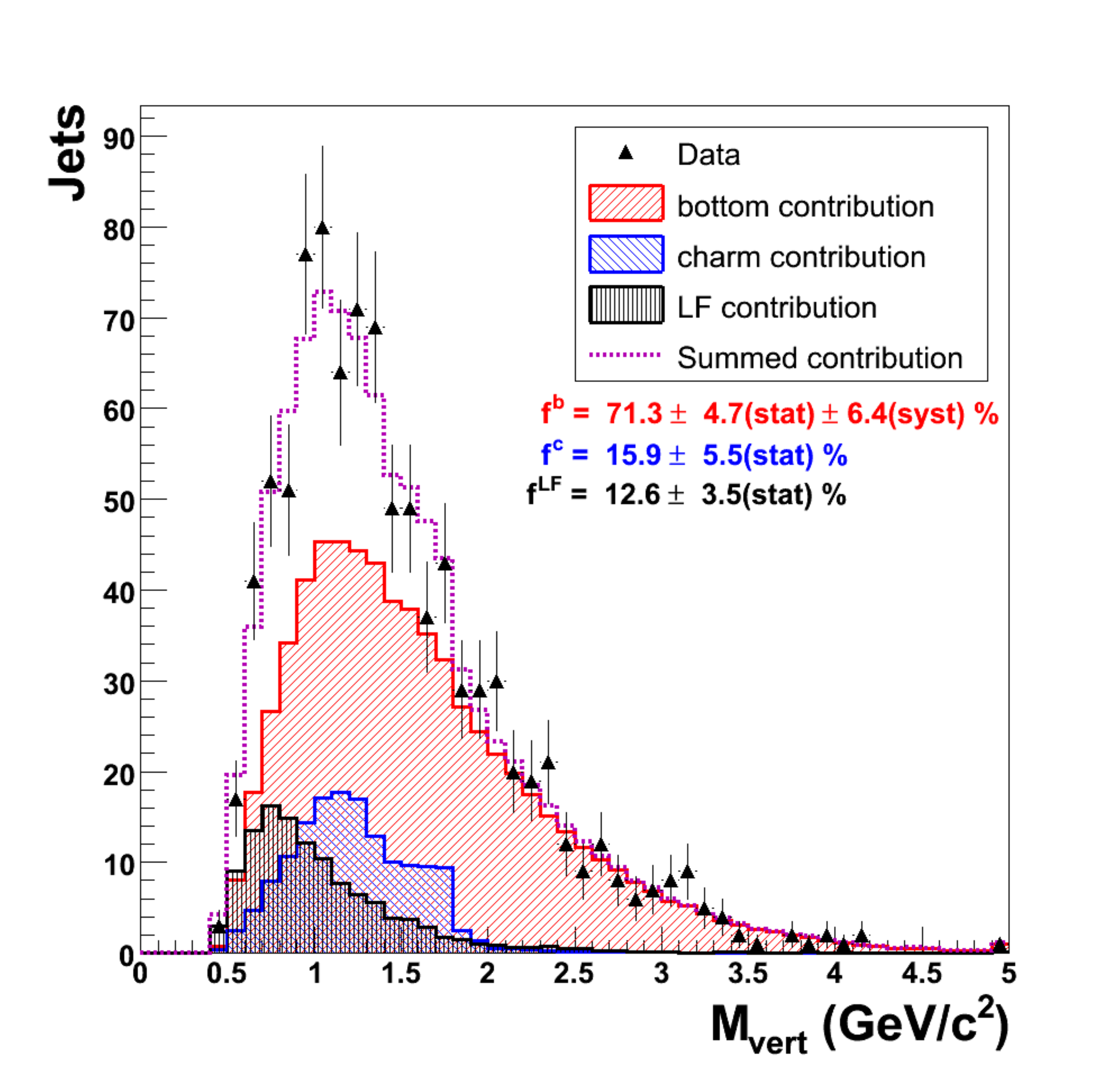}
        }
\hfill
\subfigure[ATLAS measurement of the $W+b$-jet cross section in 
the 1, 2, and 1+2 jet exclusive bins. The measurements are 
compared with NLO predictions in pQCD. 
]{%
           \label{f:ATLASWb}
           \includegraphics[width=0.46\textwidth]{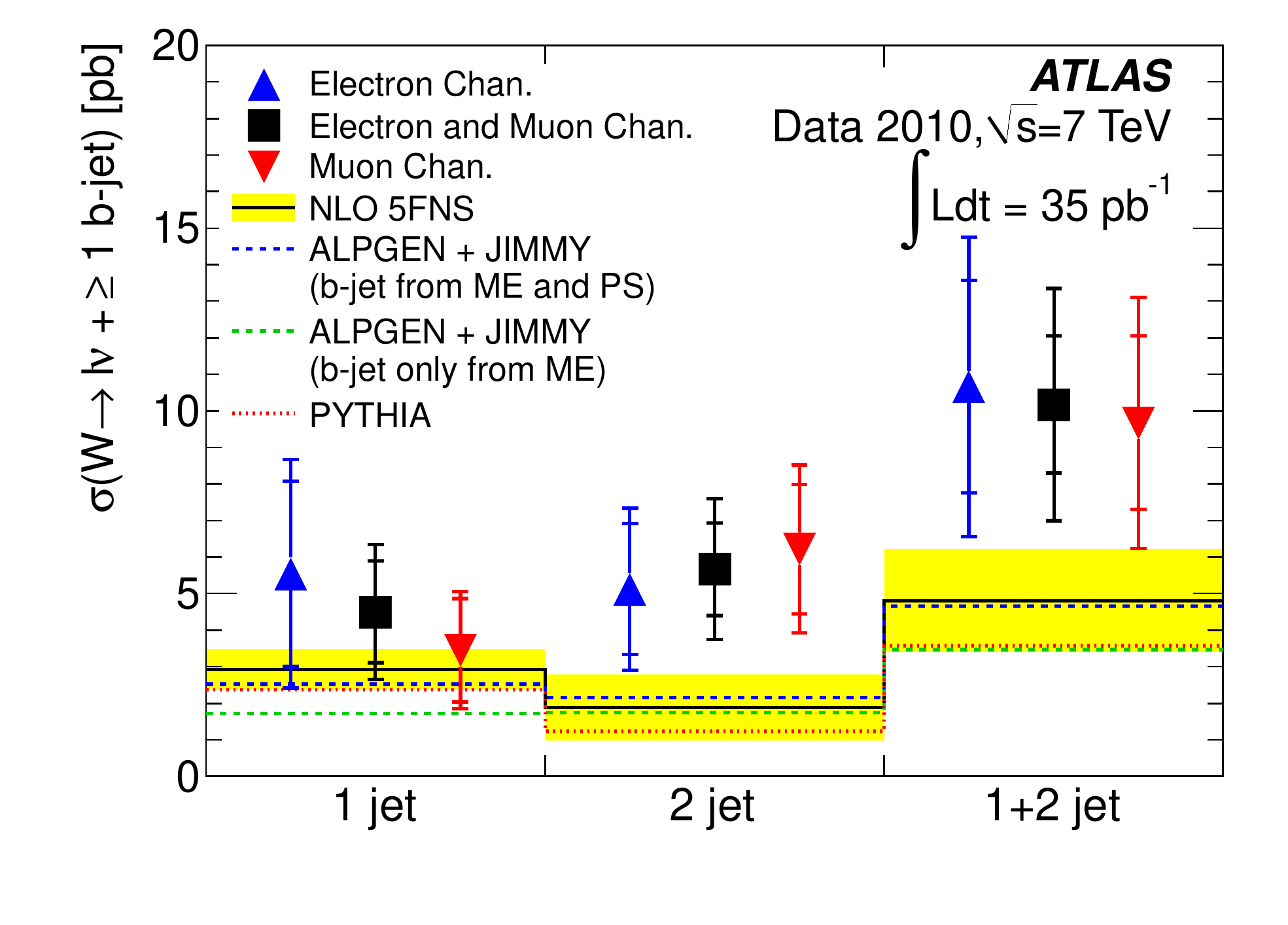}
        }
\caption[*]{CDF and ATLAS measurements of the $W+b$ production cross section.}
\end{center}
\label{f:Wb}
\end{figure}

Measurements of $W/Z$ plus heavy flavor jets were also presented by experiments
at the Tevatron and LHC.  These processes are important backgrounds to 
measurement of associated Higgs boson production and suffer from 
significant theoretical
uncertainty on their production.  An earlier measurement~\cite{CDFWb} 
of $W+b$-jet production using 1.9\,fb$^{-1}$ of Tevatron Run~2 data by the 
CDF Experiment is shown in Fig.~\ref{f:CDFWb}, where 
the exclusive $W+b$-jet cross section is determined after removal of
$c$-jet and light-flavor jet background using templates based on the 
secondary vertex mass in $b$-tagged jets.  This result yielded 
a cross section of about three standard deviations in excess of SM
predictions at NLO accuracy for events with exactly one or two jets
with $E_T>20$\,GeV and lepton from $W$ decay with similar requirement.  
A more recent measurement~\cite{ATLASWb} performed by ATLAS considers 
exclusive production of $W+b$-jet events for one, two, and one or two
identified $b$ jets.  The results shown in Fig.~\ref{f:ATLASWb} are
in good agreement with NLO predictions for the one jet exclusive case, and 
agree to within 1.5 s.d. for the other categories.  A measurement~\cite{D0Wb}
provided by the D0 Experiment following the conference finds the inclusive
$W+b$-jet production cross section to be in agreement with NLO predictions 
within scale and PDF uncertainties.

\begin{figure}[!thb]
\begin{center}
\subfigure[CMS measurement of the $p_T$ of the $b$-jet 
system in events selected to identify $Z+bb$ final states.]{%
            \label{f:CMSZb}
            \includegraphics[width=0.46\textwidth]{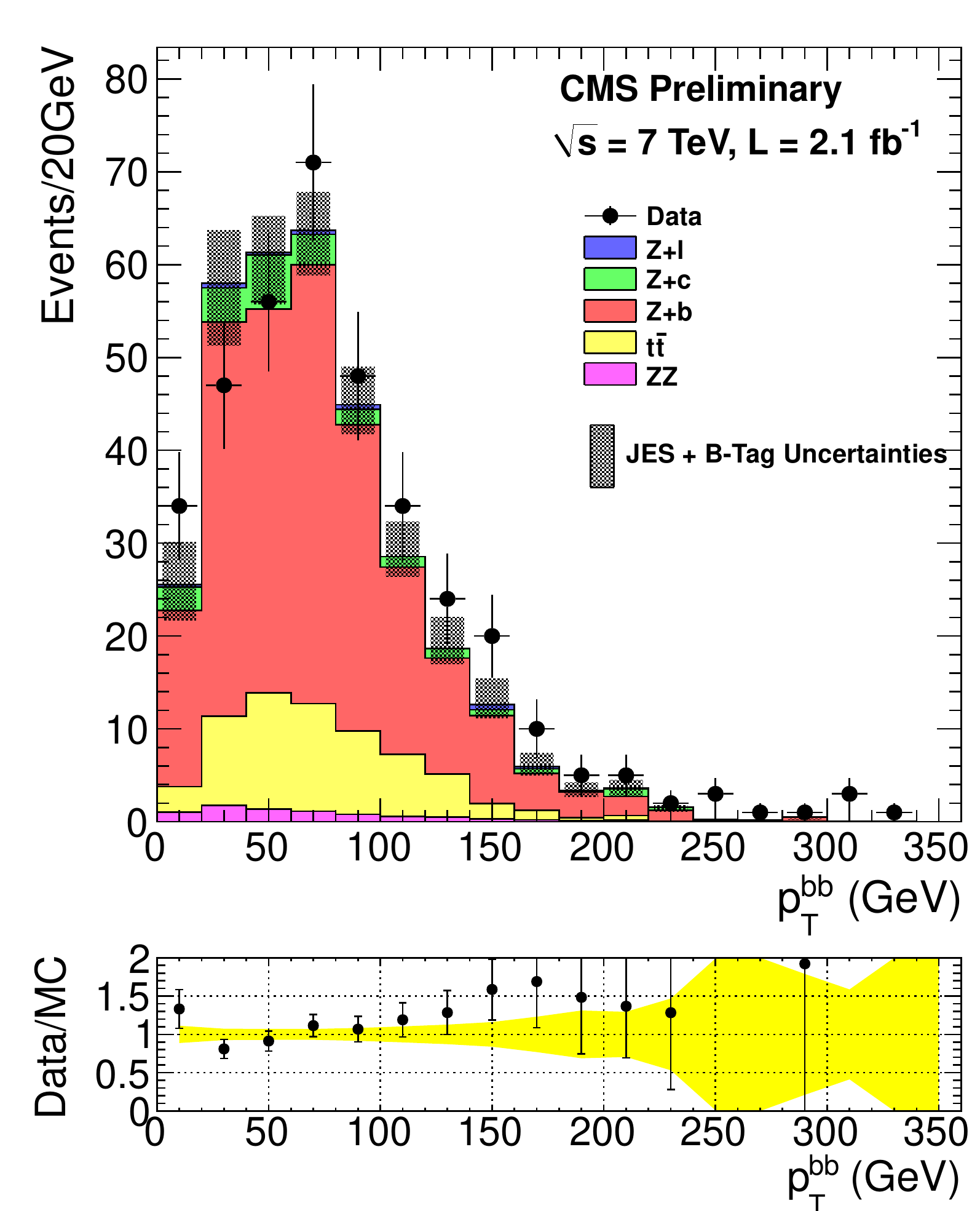}
        }
\hfill
\subfigure[D0 ratio of $\gamma+b$ differential cross section 
between data and NLO QCD predictions with uncertainties
for the rapidity region $|y^\gamma|<1.0$.]
{%
           \label{f:D0gb}
           \includegraphics[width=0.46\textwidth]{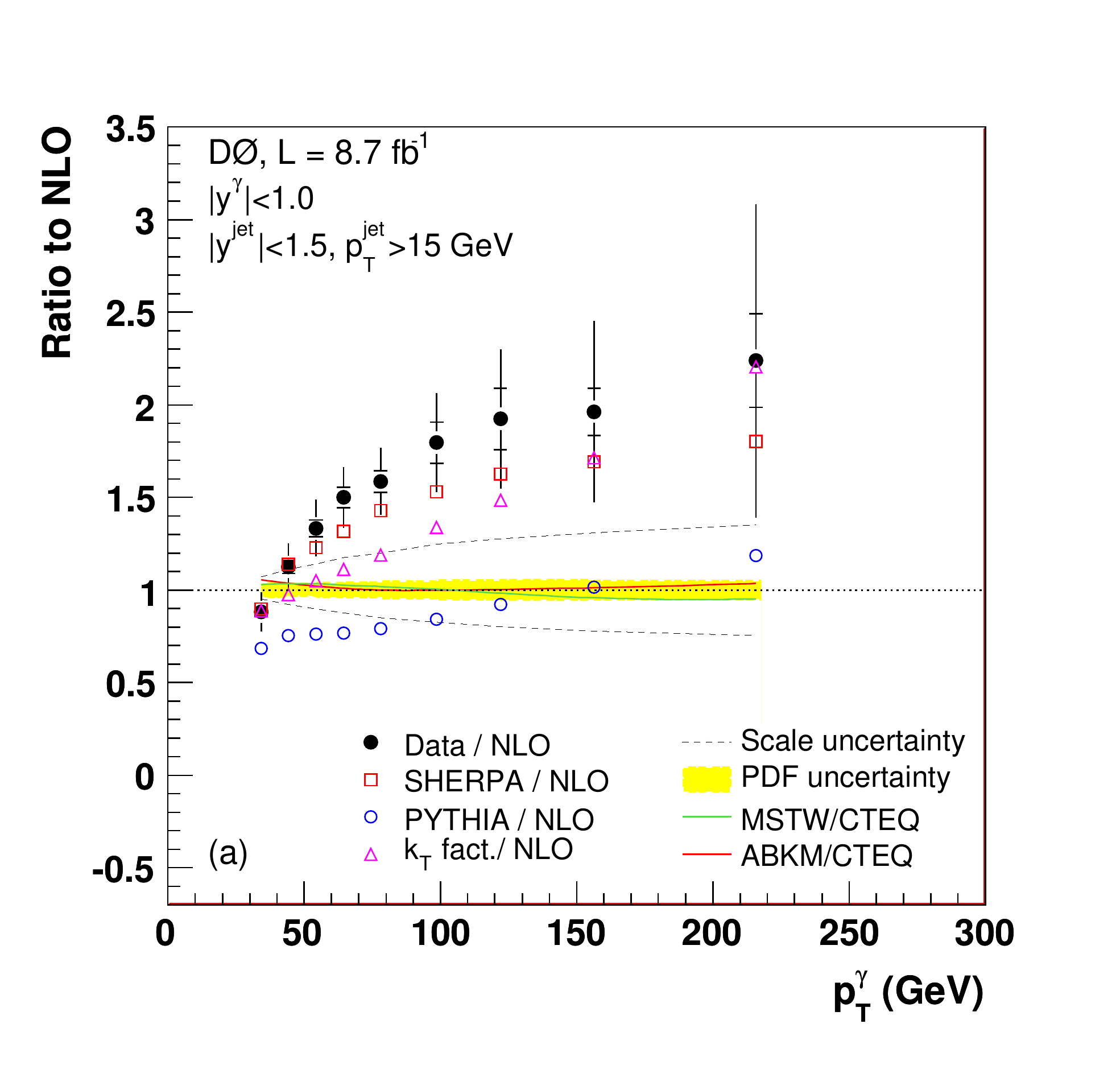}
        }
\caption[*]{Measures of the $Z+b$ and $\gamma+b$ cross sections from CMS and D0.}
\end{center}
\label{f:Zgb}
\end{figure}

The CMS Experiment provided a recent study~\cite{CMSZb} 
of the $Z/\gamma^*+b$-jet cross sections using 2.1 fb$^{-1}$ of pp collisions
at $\sqrt{s}=7$\,TeV and additional results were provided by the 
CDF Experiment~\cite{CDFZb} using the full CDF data set.
Figure~\ref{f:CMSZb} shows the $p_T$ of the $bb$ system in an event sample
selected to identify $Z+bb$ final states at CMS.  In general the CMS results 
indicate that predictions based on the {\sc MadGraph}
event generator interfaced with {\sc pythia} and normalized to 
the NNLO inclusive cross section provide a fair description of the data.

Finally a measurement~\cite{D0gb} of the differential cross section 
for photon$+b$-jet production was provided by the D0 Experiment.  The results 
are summarized in Fig.~\ref{f:D0gb} which shows the ratios of data to the 
NLO QCD calculations and different MC simulations 
to the same NLO calculations.  The observed variance with
respect to several PDF models is also shown.  The data can only  
be described by including higher order corrections into the
NLO QCD predictions, such as those currently present as additional 
real emissions in the {\sc sherpa} MC generator.  
This is an interesting measurement
to compare with future results from the LHC, since the production of
photon plus heavy flavor events at the 
Tevatron is dominated by final state gluon splitting 
$q\bar q\to\gamma g(g\to b\bar b)$, while at the LHC $bq\to b\gamma$ 
dominates for most of the experimentally accessible $p_T$ ranges.

\section{Soft QCD}

\begin{figure}[!thb]
\begin{center}
\subfigure[CMS reconstructed pseudorapidity distributions of the 
second-leading jets, after imposing a pseudorapidity gap requirement,
and compared to detector-level MC predictions with and without diffractive dijet
production.]{%
            \label{f:CMSdiff1}
            \includegraphics[width=0.46\textwidth]{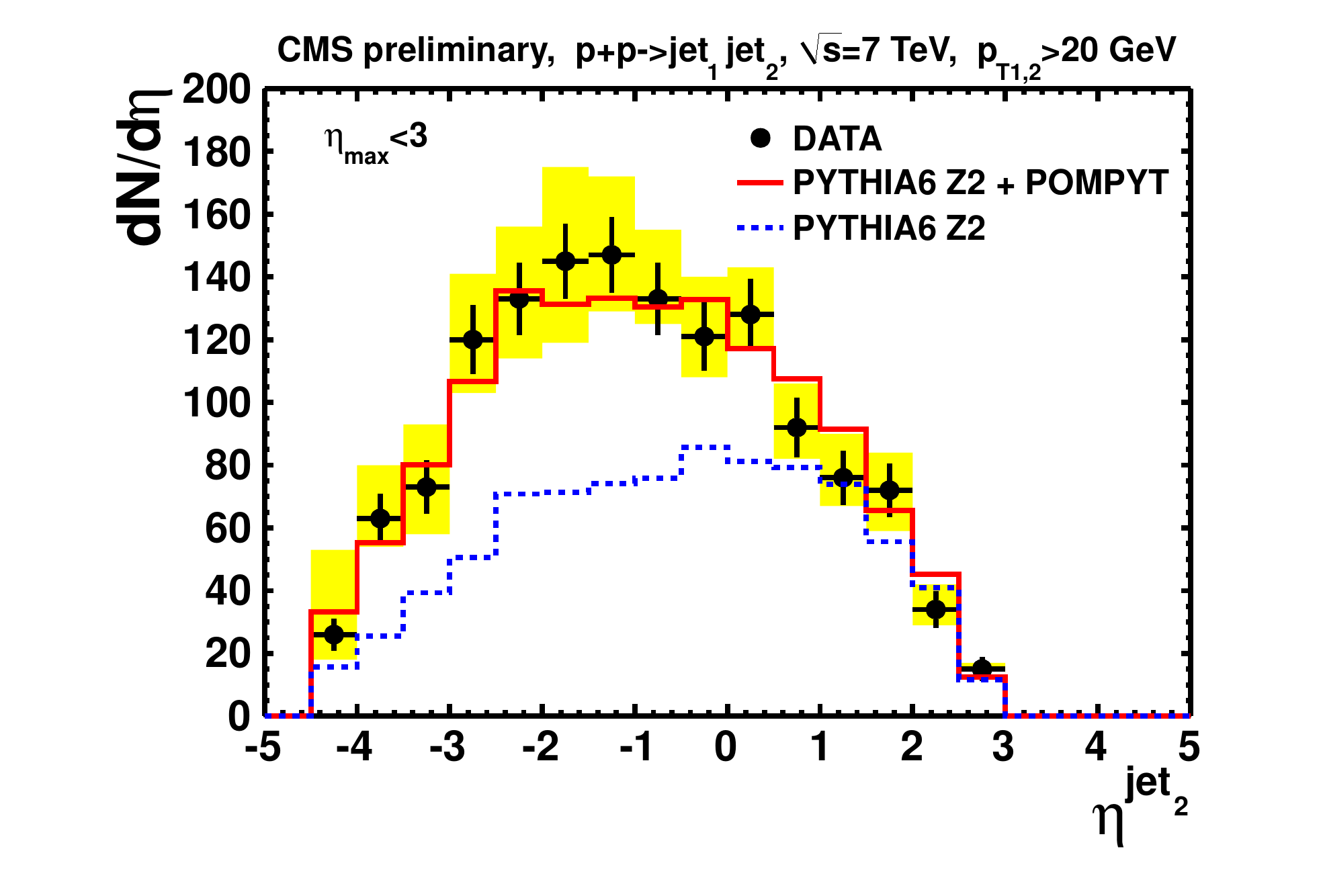}
        }%
\hfill
\subfigure[CMS reconstructed $\tilde\xi$ distribution compared to 
detector-level MC predictions with and without diffractive dijet
production.]{%
           \label{f:CMSdiff2}
           \includegraphics[width=0.46\textwidth]{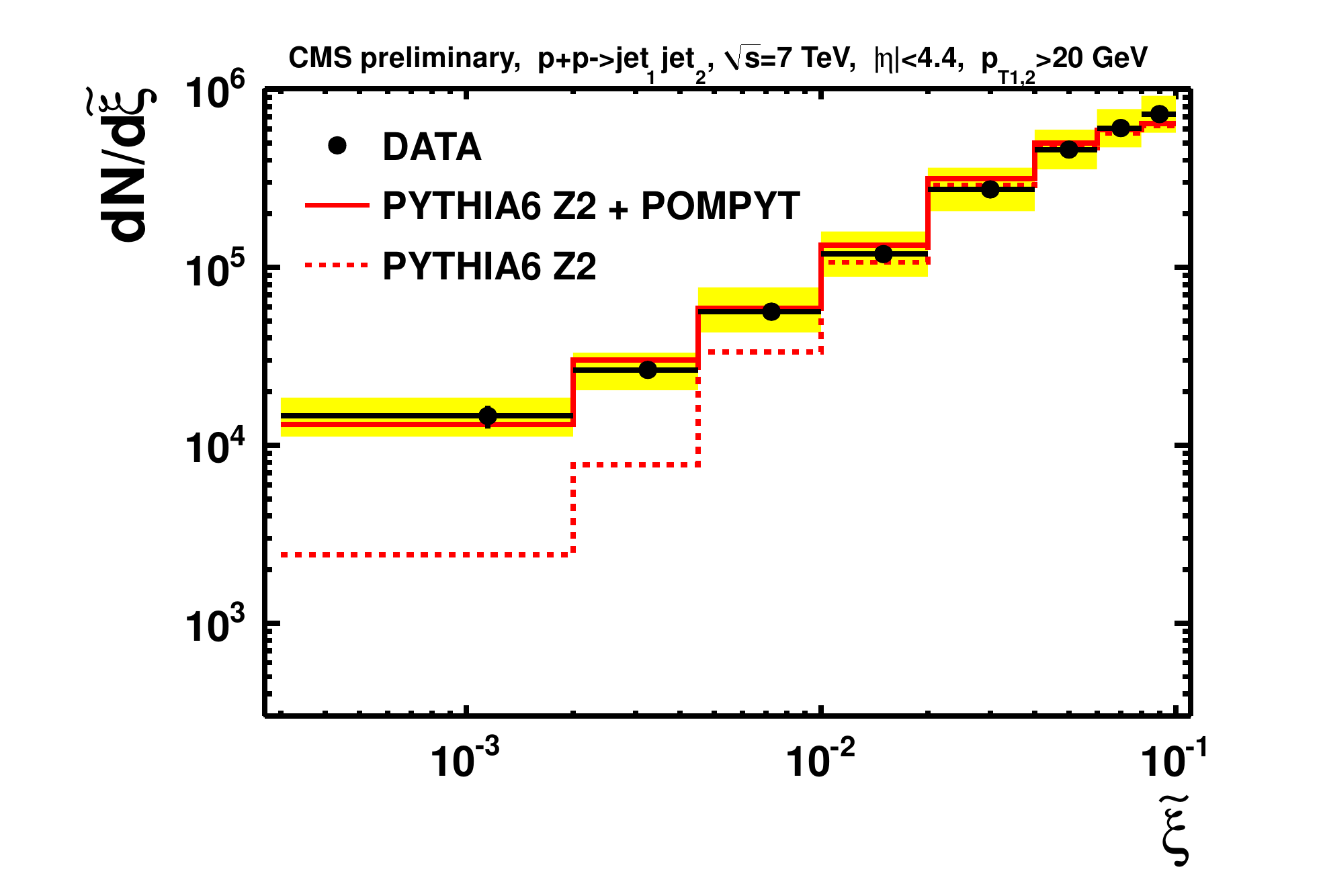}
        }
\caption[*]{CMS study of diffractive dijet production.}
\end{center}
\label{f:CMSDiff}
\end{figure}

Another area of study at the LHC and Tevatron experiments involves 
nonperturbative processes.  Reviews of measurements involving 
multiparton interactions and total cross sections are presented 
respectively by E. Dobson and N. Cartiglia in these proceedings.
Presented here is a recent measurement~\cite{CMSdiff} by the CMS experiment 
of the contribution from diffractive dijet production to the inclusive dijet
cross section.  Events are selected to contain at least two jets with
$p_T>20$\,GeV with the leading jet satisfying $|\eta^{jet}|<4.4$~\cite{eta}.
The quantity\footnote{In this study of diffractively produced events 
$M_X$ represents the mass of the jet system which is separated
from the an opposing scattered proton by a large rapidity gap} $\xi=M^2_X/s$, 
which approximates the fractional momentum loss
of the scattered proton in single diffractive events is approximated 
($\tilde\xi$) from the energies and longitudinal momenta of
all particle flow~\cite{CMSPF} objects measured in the region $|\eta|<4.9$.  
Figure~\ref{f:CMSdiff1} shows the $\eta$ distribution for the second leading 
jet after application of a requirement to enhance the diffractive 
contribution to the measured cross section. This requirement~\cite{CMSdiff}
is equivalent to imposing a pseudorapidity gap of at least 1.9 units,
enhancing the diffractive component in the data, 
and selecting events with the jets mainly in the hemisphere 
opposite to that of the gap.  A combination of {\sc PYTHIA6} with tune Z2 and
single diffractive events modeled with {\sc POMPYT}~\cite{pompyt}
is found to agree with the data reasonably well.  Figure~\ref{f:CMSdiff2} shows
the reconstructed $\tilde\xi$ distribution compared to 
detector-level MC predictions with and without the inclusion of
diffractive dijet production.

\section{Conclusions}

Remarkable progress continues in measurements concerned with jet production.
Unprecedented data sets, measurement precision, and improvements in theory 
have enabled ever more rigorous tests of standard model predictions 
and the extraction of fundamental parameters such as the strong coupling 
constant at ever increasing momentum scales.  These data inform models
for PDFs and constrain important backgrounds for rare SM and new physics 
searches.  They further test the applicability of fixed order perturbative 
calculations, parton shower MC, and other models
over an exceptionally large phase
space of final states.  As shown in this brief sampling of results, numerous 
areas require work understand the effects of higher order and nonperturbative
processes
to improve our modeling of the data and further enhance 
sensitivities to new physics processes.

I wish to thank the organizers of Physics in Collision 2012 for giving me the
opportunity to present these results and the ATLAS, CDF, CMS, D0, HERA, and ZEUS
collaborations for providing the results included in this talk.


\begin{thebibliography}{0}
\bibitem{pQCD} 
F. Aversa {\em et al.}, Phys. Rev. Lett {\bf 65}, 401 (1990); \\
W. T. Giele, E.W. N. Glover, and D. A. Kosower, 
Phys. Rev. Lett {\bf 73}, 2019 (1994); \\
D. Ellis, Z. Kunszt, and D. E. Soper, 
Phys. Rev. Lett {\bf 64}, 2121 (1990).

\bibitem{PDF}
See contribution from J. Bl\"{u}mlein in these proceedings.

\bibitem{fastjet}  M. Cacciari, G. P. Salam, G. Soyez, 
CERN-PH-TH/2011-297, arXiv:1111.6097.

\bibitem{D0Cone}  V. M. Abazov {\em et al.} (D0 Collaboration),
Phys. Rev. D {\bf 85} (2012) 052006.

\bibitem{CDFCone}
A. Abulenci {\em et al.} (CDF Collaboration),
Phys. Rev. D {\bf 74}, 071103 (2006).

\bibitem{kt}
S. Catani, Y. L. Dokshitzer, M. H. Seymour and B. R. Webber, 
Nucl. Phys. B {\bf 406}, 187 (1993); \\
S. D. Ellis and D. E. Soper, Phys. Rev. D {\bf 48}, 3160 (1993).

\bibitem{antikt}
M. Cacciari and G. P. Salam, JHEP {\bf 0804} 005 (2008).

\bibitem{D0incjets} V. Abazov {\em et al.} (D0 Collaboration),
Phys. Rev. D {\bf 85}, 052006 (2012);
V. Abazov {\em et al.} (D0 Collaboration),
Phys. Rev. Lett. {\bf 101}, 062001 (2008). 

\bibitem{CMSincdijets} S. Chatrchyan {\em et al.} (CMS Collaboration),
[arXiv:1212.6660].

\bibitem{ATLASdijet7} G. Aad {\em et al.} (ATLAS Collaboration),
Phys. Rev. D {\bf 86} (2012) 014022.

\bibitem{ATLASdijet8} B. Chapleau, presentation at 
36th International Conference for High Energy Physics, July 2012,\\
http://indico.cern.ch/conferenceDisplay.py?confId=181298.

\bibitem{nnpdf} R. D.~Ball {\it et al.}, Nucl.Phys. B {\bf 838}, 136 (2010).


\bibitem{D0RDR}  V. Abazov {\em et al.} (D0 Collaboration),
Phys. Lett. B {\bf 718}, 56 (2012).


\bibitem{CMSmulti} V. Khachatryan {\em et al.} (D0 Collaboration), 
Phys. Lett. B {\bf 699} 48-67( 2011).

\bibitem{tauC} A. Banfi, G.P. Salam, G. Zanderighi, JHEP 1006 (2010) 038.

\bibitem{pythia} T. Sjostrand, S. Mrenna and P. Z. Skands, J. High Energy
Phys. 05, 26 (2006).

\bibitem{herwig}
G.~Marchesini et al., {\it The Herwig Event Generator},
Comp. Phys. Comm. {\bf 67} (1992) 465.

\bibitem{alpgen} M. L. Mangano et al., J. High Energy Phys. 07, 1 (2003).

\bibitem{madgraph} J. Alwall, M. Herquet, F. Maltoni, O.Mattelaer, T. Stelzer,
J. High Energy Phys. {\bf 06} (2011) 1029.


\bibitem{nlojet++}
Z. Nagy, Phys. Rev. Lett. {\bf 88} (2002) 122003 [hep-ph/0110315]; 
Phys. Rev. D {\bf 68} (2003) 094002.


\bibitem{ct10} J. Gao {\em et al.}, arXiv:1302.6246.

\bibitem{Hess} A. Cooper-Sarkar and C. Gwenlan, in Proceedings of the
Workshop: HERA and the LHC, Part A, edited by A. De
Roeck and H. Jung, Geneva, Switzerland (2005), CERN-
2005-014, DESY-PROC-2005-01, arXiv:hep-ph/0601012,
see part 2, section 3.

\bibitem{D0alphas}  V. Abazov {\em et al.} (D0 Collaboration),
Phys. Rev. D {\bf 80}, 111107 (2009).

\bibitem{sherpa} T. Gleisberg et al., J. High Energy Phys. 02, 7 (2009).

\bibitem{hej} J. R. Andersen and J. M. Smillie, J. High Energy Phys. 
01, 39 (2010); J. R. Andersen and J. M. Smillie, Phys. Rev. D 81,
114021 (2010); J. R. Andersen and J. M. Smillie, Nucl. Phys. Proc.
Suppl. 205, 205 (2010);
J. R. Andersen and J. M. Smillie, J. High Energy Phys.
06, 10 (2011);
J. R. Andersen, T. Hapola and J. M. Smillie, J. High
Energy Phys. 09, 047 (2012).

\bibitem{blackhat}
C. F. Berger, et al., Phys. Rev. Lett. 102, 222001 (2009);
C. F. Berger et al., Multi-jet cross sections at NLO with
BlackHat and Sherpa, arXiv:0905.2735 [hep-ph];
C. F. Berger et al., Phys. Rev. D 80, 074036 (2009).

\bibitem{rocket}J. M. Campbell and R. K. Ellis, Phys. Rev. D 65, 113007
(2002);
J. M. Campbell, R. K. Ellis and D. L. Rainwater, Phys.
Rev. D 68, 94021 (2003).

\bibitem{mcfm} R. K. Ellis et al., J. High Energy Phys. 01, 12 (2009);
W. T. Giele and G. Zanderighi, J. High Energy Phys. 06,
38 (2008).

\bibitem{D0Wj} V. Abazov {\em et al.} (D0 Collaboration),
[arXiv:1302.6508].

\bibitem{ATLASWj} G. Aad {\em et al.} (ATLAS Collaboration),
Phys. Rev. D {\bf 85} (2012) 092002.

\bibitem{NLOEW}  A. Denner, S. Dittmaier, T. Kasprzik, A. M\"{u}ck,
JHEP {\bf 06} (2011) 069.

\bibitem{CDFZlpt} http://www-cdf.fnal.gov/physics/new/qcd/zjets10fb\_new/NLOEW.html.

\bibitem{CDFWb} T. Aaltonen {\em et al.} (CDF Collaboration),
Phys. Rev. Lett. {\bf 104} 131801 (2010).

\bibitem{ATLASWb}  G. Aad {\em et al.} (ATLAS Collaboration),
Phys. Lett. B {\bf 707} 418 (2012).

\bibitem{D0Wb} V. Abazov {\em et al.} (D0 Collaboration),
Phys. Lett. B {\bf 718} 1314 (2013).

\bibitem{CMSZb} S. Chatrchyan {\em et al.} (CMS Collaboration),
J. High Energy Phys. {\bf 06} (2012) 126.

\bibitem{CDFZb} T. Aaltonen {\em et al.} (CDF Collaboration),
http://www-cdf.fnal.gov/physics/new/qcd/zbjet2012/index.html.

\bibitem{D0gb} V. Abazov {\em et al.} (D0 Collaboration),
Phys. Lett. B {\bf 714} 32 (2012).

\bibitem{CMSdiff} S. Chatrchyan {\em et al.} (CMS Collaboration),
Phys. Rev. D {\bf 87} (2013) 012006.

\bibitem{eta}
The pseudorapidity is defined as $\eta=-\ln[\tan(\theta/2)]$, 
where $\theta$ is the polar angle with respect to one of the
beam directions.

\bibitem{CMSPF} CMS Collaboration, CMS Physics Analysis Summary
Report No. CMS-PAS-PFT-09-001, 2009.

\bibitem{pompyt} P. Bruni and G. Ingelman, “Diffractive hard scattering at ep and p anti-p colliders”, DESY
93-187, (1993).



\end{thebibliography}
\end{document}